\documentclass[amsmath,trackchanges,twocolumn,longbib,times]{aastex702}

\usepackage[T1]{fontenc}
\usepackage{graphicx}%
\usepackage{tabularx}%
\usepackage{xltabular}
\usepackage{booktabs}%
\usepackage{subcaption}
\usepackage{bm}
\usepackage{amsmath}	
\usepackage{amssymb}	
\usepackage{xcolor}
\usepackage{siunitx}
\usepackage{makecell}
\usepackage{orcidlink}
\usepackage[]{hyperref}
\usepackage{ulem}
\usepackage{threeparttable}
\usepackage{needspace}
\usepackage{makecell}
\hypersetup{colorlinks,citecolor=blue, linkcolor=blue, urlcolor=blue, breaklinks=true}

\usepackage{placeins}           

\begin{document}

\title{Size, Albedo, and Taxonomy of Asteroid (1201) Strenua from Stellar Occultation and ATLAS Photometry}

\correspondingauthor{Agus T. P. Jatmiko}

\author[orcid=0009-0000-6563-3069,sname='Jatmiko']{Agus T. P. Jatmiko}
\affiliation{Bosscha Observatory, Faculty of Mathematics and Natural Sciences, Institut Teknologi Bandung, Bandung, Indonesia}
\email[show]{agustriono.pj@itb.ac.id}  

\author[orcid=0009-0002-1279-1906,sname='Yusuf']{Muhammad Yusuf} 
\affiliation{Bosscha Observatory, Faculty of Mathematics and Natural Sciences, Institut Teknologi Bandung, Bandung, Indonesia}
\email{m.yusuf@itb.ac.id}

\author[orcid=0000-0001-7663-5801, sname='Dermawan']{Budi Dermawan}
\affiliation{Bosscha Observatory, Faculty of Mathematics and Natural Sciences, Institut Teknologi Bandung, Bandung, Indonesia}
\affiliation{Astronomy Research Group, Faculty of Mathematics and Natural Sciences, Institut Teknologi Bandung, Bandung, Indonesia}
\email{budider@itb.ac.id}

\author[orcid=0009-0007-9538-7146, sname='Perhati']{Teduh Perhati}
\affiliation{Bosscha Observatory, Faculty of Mathematics and Natural Sciences, Institut Teknologi Bandung, Bandung, Indonesia}
\affiliation{Master's Program in Astronomy, Faculty of Mathematics and Natural Sciences, Institut Teknologi Bandung, Bandung, Indonesia}
\email{teduhperhati@gmail.com}

\author[orcid=0009-0008-0518-5928, sname='Ramadhan']{Dhimaz G. Ramadhan}
\affiliation{Bosscha Observatory, Faculty of Mathematics and Natural Sciences, Institut Teknologi Bandung, Bandung, Indonesia}
\email{dhimaz.gilang@gmail.com}

\author[orcid=0009-0009-7162-0251, sname='Aprianizar']{Muhammad Z. Aprianizar}
\affiliation{Bosscha Observatory, Faculty of Mathematics and Natural Sciences, Institut Teknologi Bandung, Bandung, Indonesia}
\email{aprianizakky13@gmail.com}

\author[orcid=0009-0001-9373-6660, sname='Fatkhusalma']{Satrio Fatkhusalma}
\affiliation{Bosscha Observatory, Faculty of Mathematics and Natural Sciences, Institut Teknologi Bandung, Bandung, Indonesia}
\email{satrio.fatkhusalma@gmail.com}

\author[orcid=0009-0000-0055-9021, sname='Mandey']{Denny Mandey}
\affiliation{Bosscha Observatory, Faculty of Mathematics and Natural Sciences, Institut Teknologi Bandung, Bandung, Indonesia}
\email{mandey.de@gmail.com}

\author[]{STRENUA-26 Contributors}
\affiliation{A full list of the STRENUA-26 Contributors is provided in Appendix~\ref{sec:appendix_contributors}.}
\email{budider@itb.ac.id}


\begin{abstract}
	We determine the size, geometric albedo, and taxonomic classification of asteroid (1201) Strenua by combining a multi-chord stellar occultation observed on April 26, 2026 with ATLAS $c$- and $o$-band photometry. A coordinated campaign deployed 47 observing sites across Indonesia along the predicted shadow path, returning four positive chord detections. Fitting an ellipse to the asteroid's instantaneous cross-section within a Bayesian forward-modeling framework yields apparent semi-axes of $a' = 27.49^{+3.90}_{-1.65}$~km and $b' = 14.33^{+2.37}_{-2.69}$~km at a position angle of $95.3^{+4.9}_{-2.7}$~deg, corresponding to an area-equivalent diameter of $D_\mathrm{eq} = 39.86^{+2.06}_{-1.66}$~km. Independently, modeling the ATLAS $c$- and $o$-band photometry with the three-parameter $H, G_1, G_2$ phase-function system gives $H_c = 11.50 \pm 0.05$~mag, $G_{1,c} = 0.81 \pm 0.08$, and $G_{2,c} = 0.04 \pm 0.02$, alongside $H_o = 11.29^{+0.02}_{-0.03}$~mag, $G_{1,o} = 0.94^{+0.03}_{-0.05}$, and $G_{2,o} = 0.01 \pm 0.01$. Converting these band-specific values to the standard $V$ band provides an absolute magnitude of $H_\mathrm{V} = 11.46 \pm 0.05$~mag, which, combined with the occultation-derived diameter, establishes a geometric albedo of $p_\mathrm{V} = 0.029 \pm 0.003$. Together with the $G_1$ and $G_2$ phase parameters, this low albedo places Strenua among the darkest primitive asteroids and securely within the P-class taxonomic cluster, resolving the ambiguity of its legacy X-complex classification.
\end{abstract}



\section{Introduction}
\label{sec:intro}

Asteroids are remnants of the primordial protoplanetary disk and thus preserve a record of the earliest stages of Solar System formation that has been largely erased from the terrestrial planets by geological reprocessing (\citealp{Bottke2002}; \citealp{DeMeo2014}). Their size-frequency and compositional distributions encode the dynamical history of the Solar System, including giant planet migration \citep{Walsh2011}, while the continuous delivery of main-belt material to Earth-crossing orbits via resonances and the Yarkovsky effect sustains the near-Earth object (NEO) population \citep{Bottke2006}. The hazard posed by asteroid impacts, from the Cretaceous--Paleogene extinction event \citep{Alvarez1980} to the 2013 Chelyabinsk superbolide \citep{Brown2013}, further motivates the physical characterization of these bodies. Obtaining accurate sizes, shapes, rotational states, and compositional classifications for a statistically representative sample of the asteroid population is therefore a core objective of planetary science.

Several complementary techniques have been developed to retrieve asteroid physical parameters. Spacecraft flybys and rendezvous missions (e.g., NEAR-Shoemaker, Hayabusa, OSIRIS-REx) yield definitive sizes and shapes (\citealp{Thomas1996}; \citealp{Lauretta2019}) but are necessarily limited to a handful of targets. Radar ranging provides detailed shape models and spin states for close-approaching bodies (\citealp{Ostro2002}; \citealp{Benner2015}), but sensitivity drops as the fourth power of distance, restricting coverage to the orbitally biased NEO population. Adaptive optics (AO) resolves the largest main-belt objects (\citealp{Marchis2006}; \citealp{Viikinkoski2018}), yet its angular resolution threshold of $\sim$50--100 mas excludes the vast majority of main-belt asteroids at typical heliocentric distances. Thermal infrared radiometry, via IRAS \citep{Tedesco2002} and WISE/NEOWISE \citep{Masiero2011}, has provided diameters and albedos for tens of thousands of asteroids, but systematic uncertainties in radiometric diameters can reach 10--20\% due to model-dependent assumptions on the thermal parameter and beaming function \citep{Harris2002}. Photometric lightcurve inversion (\citealp{Kaasalainen2001a};\citealp{Kaasalainen2001b}) yields rotation periods and convex shape models for hundreds of objects archived in DAMIT \citep{Durech2010}, but is inherently scale-free: absolute sizes cannot be recovered without an independent albedo, and the method cannot resolve concave features or uniquely determine the spin-pole longitude.

Stellar occultations, in which an asteroid's shadow traverses Earth and momentarily extinguishes a background star, overcome many of the above limitations. Multiple observers distributed across the shadow path independently time the star's disappearance and reappearance; each timing pair defines a chord across the asteroid's projected silhouette, and the chord ensemble reconstructs that silhouette in a model-independent way (\citealp{Millis1989}; \citealp{Herald2024}). No assumptions about albedo, thermal properties, or convexity are required, and the absolute linear scale is measured directly. When combined with a lightcurve-inversion shape model, the occultation silhouette provides a direct size calibration with typical uncertainties of 5--10\% (\citealp{Durech2011}; \citealp{Hanus2017}). Multi-chord events can additionally detect satellites and ring systems through secondary flux drops (\citealp{BragaRibas2014}), and the resulting astrometry (particularly when referenced to the Gaia catalogue) rivals or exceeds the precision of classical CCD astrometry, improving orbital solutions significantly (\citealp{Spoto2017}; \citealp{Ferreira2022}). The publication of Gaia Data Releases 2 and 3 has reduced shadow-path uncertainties from hundreds of kilometres to a few kilometres for main-belt targets, enabling reliable multi-station campaigns for objects well below the AO resolution limit \citep{Ferreira2022}. The technique is moreover accessible to coordinated networks of modest ground-based instruments, making it uniquely scalable for population-level characterization \citep{Herald2020}.

(1201) Strenua was discovered on September 14, 1931 by K. W. Reinmuth at Heidelberg and orbits in the inner main belt. Historically, under the Bus--DeMeo taxonomy, it was classified as an X-type asteroid (\citealp{Bus2002}), a spectroscopically degenerate complex characterized by a featureless, moderately sloped visible reflectance spectrum. This legacy X-complex encompassed three compositionally distinct subtypes separable only by geometric albedo $p_\mathrm{V}$: enstatite-rich E-types ($p_\mathrm{V} > 0.30$), metallic M-types ($0.10 < p_\mathrm{V} < 0.30$) interpreted as exposed iron-nickel cores of differentiated parent bodies, and dark P-types ($p_\mathrm{V} < 0.10$) with possible organic or carbonaceous surface compositions (\citealp{Tholen1989}; \citealp{Fornasier2011}). Recently, the probabilistic taxonomy introduced by \citet{Mahlke2022} has resolved this historical degeneracy by directly reincorporating visual albedo as a fundamental classification observable. In this modernized framework, the generic X-complex is effectively dissolved: P-types are formally reassigned to the carbonaceous C-complex, while the remaining metallic and silicaceous bodies form a newly defined M-complex. Because albedo cannot be inferred from a featureless reflectance spectrum alone, an independently measured diameter remains essential to resolve the object's true nature. The combination of a direct size with the absolute magnitude $H_\mathrm{V}$ immediately yields $p_\mathrm{V}$, enabling firm taxonomic assignment under the Mahlke framework and constraining the body's thermal history and compositional provenance. Despite its classification, Strenua has received minimal physical characterization. Its diameter has been estimated from thermal infrared radiometry, first from single-epoch IRAS observations \citep{Tedesco2002} and subsequently from WISE/NEOWISE photometry \citep{Masiero2011}, both subject to the systematic uncertainties described above; no published shape model, pole solution, or occultation-derived size exists for this object.

These lingering ambiguities motivate the present study, which addresses three specific gaps. (i) The radiometric diameter of Strenua (derived independently from IRAS and WISE/NEOWISE thermal photometry) carries systematic uncertainties of up to 20\% for objects in the legacy X-complex, and no model-independent geometric measurement has yet been published to confirm or refine either estimate. (ii) Strenua's definitive placement within the modernized taxonomy (distinguishing, for instance, a carbonaceous P-type from a metallic M-type) cannot be achieved without a precise, independent albedo, which in turn requires a diameter free of the thermal-model systematics described above. (iii) Beyond the simple sphere assumed by thermal models, Strenua's shape (its projected elongation and, by extension, its pole orientation) remains entirely unconstrained, reflecting a broader paucity of fundamental physical data (shape, pole, bulk density) for intermediate-sized, featureless inner-main-belt asteroids.

We address these gaps through an extensive multi-station campaign observing the occultation of HIP 35933 by (1201) Strenua on April 26, 2026, leveraging both professional facilities and a massive citizen science outreach effort. Organized by Bosscha Observatory and deploying 47 coordinated stations across the predicted path, the campaign successfully secured a multi-chord cross-sectional profile of the asteroid. This direct geometric measurement yields the first independent equivalent diameter and projected elongation of Strenua, providing an improved geometric albedo and resolving its historical taxonomic ambiguity. Consequently, this work establishes a definitive size anchor for future 3D shape modeling and contributes to the growing catalog of rigorous, occultation-derived asteroid properties in the Gaia era.

The remainder of this paper is organized as follows. Section~\ref{sec:prediction} describes the prediction methodology and the observing campaign. Section~\ref{sec:data} presents the data reduction procedures and the timing determinations from the observational data. Section~\ref{sec:chord} details the multi-chord fitting process and the derived cross-sectional silhouette. Section~\ref{sec:albedo} discusses the updated physical properties of Strenua, focusing on its geometric albedo and its resulting classification within the modernized taxonomic framework. Finally, Section~\ref{sec:conclusions} summarizes the main conclusions of this study.


\section{Prediction and Observations}\label{sec:prediction}
\subsection{Occultation Prediction}

\begin{figure*}[ht]
	\centering
	\includegraphics[height=0.3\textheight]{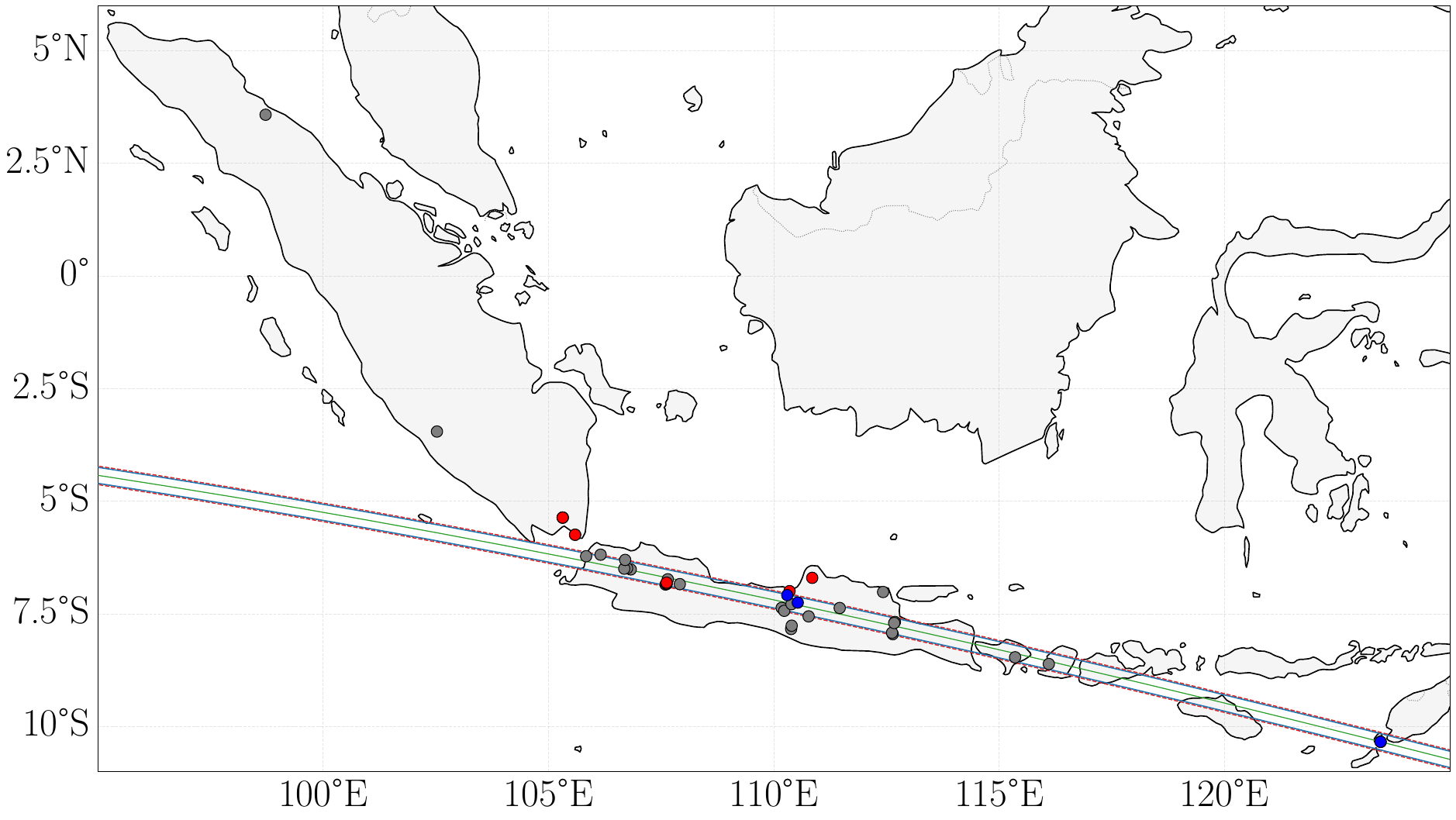}
	\caption{Observation layout and reconstructed shadow path for the April 26, 2026 occultation by (1201) Strenua. The solid blue lines trace the nominal boundaries of the projected shadow, flanked by red dashed lines illustrating the $1\sigma$ path uncertainties. Participating observatories (cf. Table~\ref{tab:list_all_stations}) are indicated by colored circles: blue for successful detections, red for negative results, and gray for stations yielding no data (NDR).}\label{fig:pred_map}
\end{figure*}

The occultation geometry was predicted using the Occult4 software\footnote{\url{http://www.lunar-occultations.com/iota/occult4.htm}} \citep{Herald2012} by projecting the asteroid's ephemerides and Gaia DR3 stellar astrometry onto the Earth's reference geoid. Initial calculations were performed in December 2025, and a final astrometric update was applied in March 2026 to propagate the cross-track uncertainties derived from the combined stellar and ephemeris errors. (1201) Strenua was predicted to occult the bright ($V = 6.4$) star HIP~35933, yielding an expected magnitude drop of $\Delta m = 10.6$. Adopting Bosscha Observatory as the reference site, the predicted topocentric mid-time was 12:41.8 UTC, with a maximum estimated occultation duration of 1.52~s (cf. Table~\ref{tab:occ_circumstances}). The resulting ground track, illustrated alongside the geographical distribution of the observing network in Fig.~\ref{fig:pred_map}, indicated a highly favorable path traversing densely populated regions across Indonesia. Comprehensive details regarding the station placements and the observing strategy are discussed in the subsequent two subsections.

%
%

\begin{table*}
	\centering
	\caption{Occultation circumstances\label{tab:occ_circumstances}}
	\begin{tabular}{ll}
		\hline\hline
		\multicolumn{2}{l}{\textit{\textbf{Occultation Prediction}}} \\
		Reference Site & Bosscha Observatory \\
		Longitude of Reference Site & $+107^\circ 37' 0.22''$ \\
		Latitude of Reference Site & $-6^\circ 49' 28.71''$ \\
		Estimated Event Date and Time & April 26, 2026 12:41.8 UTC \\
		Estimated Maximum Duration & 1.52 s \\
		\hline
		\multicolumn{2}{l}{\textit{\textbf{Star Information}}} \\
		Designation & HIP 35933 \\
		& HD 58050 \\
		& Gaia DR3 3167149432858683776 \\
		$V$ mag & 6.4 \\
		Magnitude drop ($V$ mag) & 10.6 \\
		\hline
		\multicolumn{2}{l}{\textit{\textbf{Asteroid Information and Orbital Elements}}\tablenotemark{a}} \\
		Asteroid Designation & (1201) Strenua \\
		Epoch (TDB) & JD 2461200.5 (09.0 Jun 2026) \\
		Semi-major axis, $a$ (AU) & $2.6997857424 \pm 8.6 \times 10^{-10}$ \\
		Eccentricity, $e$ & $0.0403402400 \pm 6.1 \times 10^{-10}$ \\
		Inclination, $i$ (deg) & $6.99830407 \pm 3.5 \times 10^{-8}$ \\
		Longitude of ascending node, $\Omega$ (deg) & $202.9334653 \pm 4.4 \times 10^{-7}$ \\
		Argument of perihelion, $\omega$ (deg) & $168.2562275 \pm 1.1 \times 10^{-6}$ \\
		Mean anomaly, $M$ (deg) & $125.5117936 \pm 1.2 \times 10^{-6}$ \\
		\hline
	\end{tabular}
	\tablenotetext{a}{\url{https://ssd.jpl.nasa.gov/tools/sbdb_lookup.html\#/}}
\end{table*}

\subsection{The Observing Campaign}\label{subsec:campaign}

The occultation was observed through a nationwide campaign organized by Bosscha Observatory. The campaign was built around four goals: (i) to coordinate a national network for asteroid occultation observations, (ii) to gather scientific data from multiple stations distributed along the predicted shadow path, (iii) to bring astronomy communities and amateur observers into active research, and (iv) to foster national collaboration around rare astronomical events.

Preparing such a heterogeneous network of observers required considerable groundwork. In the months before the event, Bosscha Observatory held an online colloquium, followed by a hands-on preparatory workshop and a public lecture series on the science of asteroid occultations. Contributors were supplied with a full set of technical materials, including the predicted shadow path and timing, observing guides, camera configuration instructions, a walkthrough for NTP-based time synchronization, and a detailed data submission protocol. The campaign also provided a dedicated web portal for registration, coordination, and data submission \footnote{\url{https://bosscha.itb.ac.id/id/okultasi2026/okultasi-strenua/}}. A central part of this preparation was a shared observing strategy aimed at optimizing the trade-off between achieving the highest possible signal-to-noise ratio (SNR) and maintaining temporal resolution sufficient to resolve the expected occultation duration. Because the event lasted barely more than a second, a coarse exposure time would smear ingress and egress over too few frames, while a very short exposure time would reduce the SNR to the point that the occultation would be undetectable. The final observing strategy was to use a minimum cadence of 5~Hz with unfiltered imaging, which was expected to yield a minimum SNR of $\sim$10 for the target star.

In total, the campaign mobilized 47 stations across Indonesia through a broad professional-amateur (Pro-Am) collaboration. The participating contributors ranged from university observatories and national research agencies to amateur astronomy clubs, high school science clubs, content creators, and individual observers. Given this diversity, instrumental configurations were highly heterogeneous. Well-equipped professional observatories and dedicated mobile setups primarily acquired unfiltered (clear) data to maximize the flux collected from the target star. Complementing this core scientific network, the campaign also integrated a citizen-science and educational outreach component. Amateur volunteers and high school students deployed additional mobile stations equipped with consumer color CMOS detectors, commercial DSLR cameras, and smartphone sensors. While these latter stations were primarily deployed for STEM (Science, Technology, Engineering, and Mathematics) capacity building rather than rigorous photometric acquisition, their participation significantly expanded the geographical footprint of the campaign and successfully demonstrated the feasibility of using asteroid occultations for public astronomical education. 

\subsection{Station Placement and Inter-Station Coordination}\label{subsec:st_placements}

Beyond coordinating the wider network, Bosscha Observatory fielded four of its own stations, positioned deliberately to fill gaps in chord coverage both along and across the predicted shadow path. The main observatory itself (BOSSCHA-1) in Lembang was complemented by two additional stations (BOSSCHA-3 and BOSSCHA-4) spread across the northern part of Lembang, and by a fourth station (BOSSCHA-2) placed on the center line. The mobile station BOSSCHA-2, located in Kupang, East Nusa Tenggara, played a particularly strategic role, and its placement rested on two considerations. The first was geometric: its position on the center line of the predicted shadow path. The second was meteorological: Kupang is located in a region with historically favorable weather conditions during the April-May period, in contrast to the more humid and cloud-prone western islands of Indonesia. This strategic placement of BOSSCHA-2 was crucial for ensuring a positive chord detection, especially given the high likelihood of cloud cover along the western islands during the event. These four stations were arranged to cross the shadow path at slightly different latitudes, thereby adding independent chords. The same strategy was applied to the wider network, with stations distributed to maximize the probability of positive detections while also providing broad cross-track coverage through a central communication channel maintained by Bosscha Observatory. As a result, several institutions with the resources to deploy multiple stations within their home regions increased local chord density (see Table~\ref{tab:list_all_stations}). The Astronomical Observatory of Institut Teknologi Sumatera (OAIL ITERA), for instance, operated six stations across Lampung Province. The Planetarium and Observatory of UIN Walisongo Semarang operated two stations, and the UAD Center for Astronomy Studies operated three. Clustering stations in this way provided redundancy when weather or technical faults affected some sites and, when all stations succeeded, yielded closely spaced chords that resolved the shadow path with high precision.

To ease logistical coordination across the distributed network, a synchronized observing strategy was implemented; all stations were instructed to commence continuous data acquisition at 12:37 UTC for a nominal duration of ten minutes. Of the participating sites (see Table~\ref{tab:list_all_stations} and Fig.~\ref{fig:pred_map}), four successfully recorded positive occultation chords (BOSSCHA-2, ZUL, PL-2, UINWS-2), while nine reported negative detections. It should be noted that BOSSCHA-2 and ZUL, although operated as independent stations, were deployed only a few metres apart at the same site in Kupang. The remaining majority of the network was unable to acquire usable data, primarily because of adverse weather conditions or technical problems. An additional seven sites did not submit their observational logs and are consequently excluded from the final contributor list.

\subsection{Time Synchronization Validation}

The heterogeneity of the network extended to the timing systems themselves, and this had a direct impact on the data. Of the four positive stations, only the Kupang setup (BOSSCHA-2) used hardware-level GPS timestamping, while the remaining three relied on Network Time Protocol (NTP) synchronization over the internet. Because the two schemes occupy different regimes of timing fidelity, and because the occultation event was barely longer than a second, we could not simply assume that every station's timestamps were equally trustworthy. We therefore carried out a dedicated experiment to quantify the timing behavior of each configuration before folding the chords into the geometric fit.

The validation was performed using NEXTA (New EXposure Timing Analyser), a GNSS-based timing reference introduced by \citet{Kaminski2023} that reaches sub-millisecond precision. The device is a strip of 20 LEDs driven directly by a GNSS receiver, arranged in five sections that each encode one fractional digit of the UTC second (1s, 0.1s, 0.01s, 0.001s, 0.0001s). Each section comprises four LEDs that encode digits 0-9 in a fixed pattern, so that by imaging the LED panel with the camera under test, the true UTC time can be read directly from the LED pattern. The device is designed to be imaged by a camera under test; by comparing the true UTC time encoded in the LED pattern with the timestamp the camera records for the same frame, any internal bias or latency in the acquisition chain can be exposed. This makes it well suited to our purpose, since it measures the end-to-end timing of a real camera and software configuration rather than the nominal accuracy of the clock alone. To automate the readout, we developed decoding software that locates all 20 LED positions from an integration image (all LEDs illuminated) and translates the on/off pattern of each section into its fractional-second value following the NEXTA encoding scheme.

The NTP-synchronized stations disciplined their system clocks using the Meinberg NTP Client\footnote{\url{https://www.meinbergglobal.com/english/sw/ntp.htm}}, a widely used implementation of the protocol. To give the client time to converge and stabilize, we started the NTP service at least two hours before the timing tests (and, in the field, before the event), so that the measured offsets reflect the client's stable, converged state rather than transient behavior during initial synchronization. NTP requires an extended settling period during which it repeatedly polls its time servers and gradually steers the local clock, rather than correcting in a single step.

\begin{figure}[ht]
	\centering
	\centering
	\includegraphics[height=0.22\textheight]{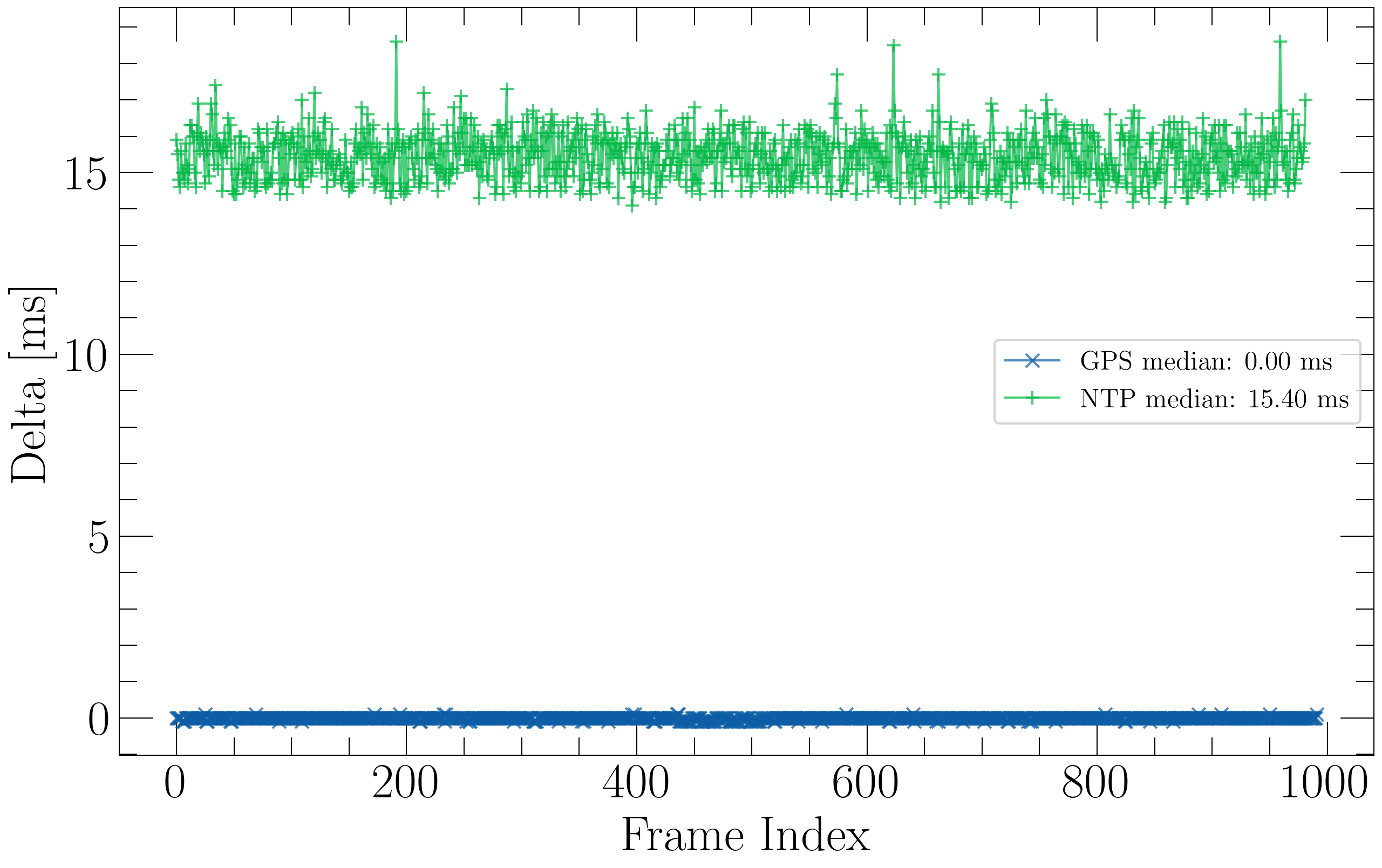}
	\caption{Per-frame timing offsets relative to true UTC, measured with NEXTA. Each point is the difference between the camera timestamp and the true UTC decoded from NEXTA LEDs, plotted against frame index over 1000 consecutive frames.}
	\label{fig:nexta_timing}
\end{figure}

Using NEXTA as a common reference, we tested two timing regimes representative of the network: the QHY174M-GPS camera with GPS active, and the same camera with GPS disabled, relying instead on an NTP-synchronized system clock. With GPS active, the difference between the camera timestamp recorded in the \texttt{DATE-AVG} header field, equivalent to \texttt{GPS\_ST} seconds plus \texttt{GPS\_SU} microseconds, and the true UTC read from NEXTA showed a median of $0.00$~ms with a scatter of $\sigma = 0.03$~ms over a thousand frames. With GPS disabled, the same camera revealed a systematic bias with a median of $\approx 15.4$~ms and a scatter of $\sigma = 0.7$~ms relative to UTC (Fig.~\ref{fig:nexta_timing}).

The tests were performed with the minimum exposure time available on the camera in 16-bit full-frame mode, 0.0355~ms, chosen to resolve the fastest blinking LED section of NEXTA. It is worth emphasizing that the residual scatter seen in the GPS-active case is not a genuine timing error of the camera, but an artifact of finite exposure time. NEXTA's highest-resolution section switches at $0.1$~ms intervals, so an exposure that is not short enough to freeze a single LED state will occasionally straddle a transition, catching the LED mid-change and yielding a $\pm 0.1$~ms ambiguity in the decoded digit. The apparent scatter is therefore a consequence of the exposure being comparable to the LED switching interval, rather than a true timing error.

These results provided a physical basis for weighting the stations differently in the multi-chord fit (Section \ref{sec:chord}). Rather than treating all chords equally, we assigned the GPS-timestamped data a statistical weight ten times higher than that of the NTP-synchronized data, reflecting the roughly two orders-of-magnitude difference in their timing fidelity. This weighting scheme was implemented in the Bayesian fitting framework, ensuring that the more reliable GPS-based chord had a proportionally greater influence on the derived geometric parameters of Strenua.

Using the same NEXTA reference, we also measured the inter-frame dead time of the camera and found it to be $\sim 90 \mu$s. This value is negligible compared with the exposure times used at the stations, so it introduces no significant timing gap in the light curves and can therefore be safely omitted from the analysis.

\subsection{Data and Light-Curve Extraction}
Every station recorded its observations as a sequence of FITS images rather than compressed video, preserving the full bit depth and the per-frame metadata required for precise photometry and timing. Following the event, the raw FITS sequences from all stations were transmitted to Bosscha Observatory, where they were collected and archived centrally.

While each contributor team was encouraged to reduce its own recordings and submit the resulting light curve to Bosscha Observatory, both as a hands-on exercise and as an independent cross-check, the final reduction used for the scientific analysis was performed centrally by a single team following one uniform pipeline. This choice was deliberate: it ensured that all light curves were reduced consistently and avoided systematic differences between stations caused by variations in reduction methodology.

The FITS sequences were extracted using PyMovie \citep{Anderson2019}, an open-source occultation aperture photometry program. Its key feature is an adaptive measurement mask: rather than using a fixed circular aperture, it builds the mask from the star image itself. Pixels are included when they rise above one standard deviation of the local background. The mask therefore conforms to the actual shape of the stellar profile. For each recording, aperture photometry was performed on the target star together with one or more nearby comparison stars. The comparison stars act as a reference for the target flux. Dividing by their combined signal removes residual atmospheric and transparency variations, thereby isolating the occultation signal from slower systematic trends.

\begin{figure}[t]
	\centering
	\includegraphics[height=0.3\textheight]{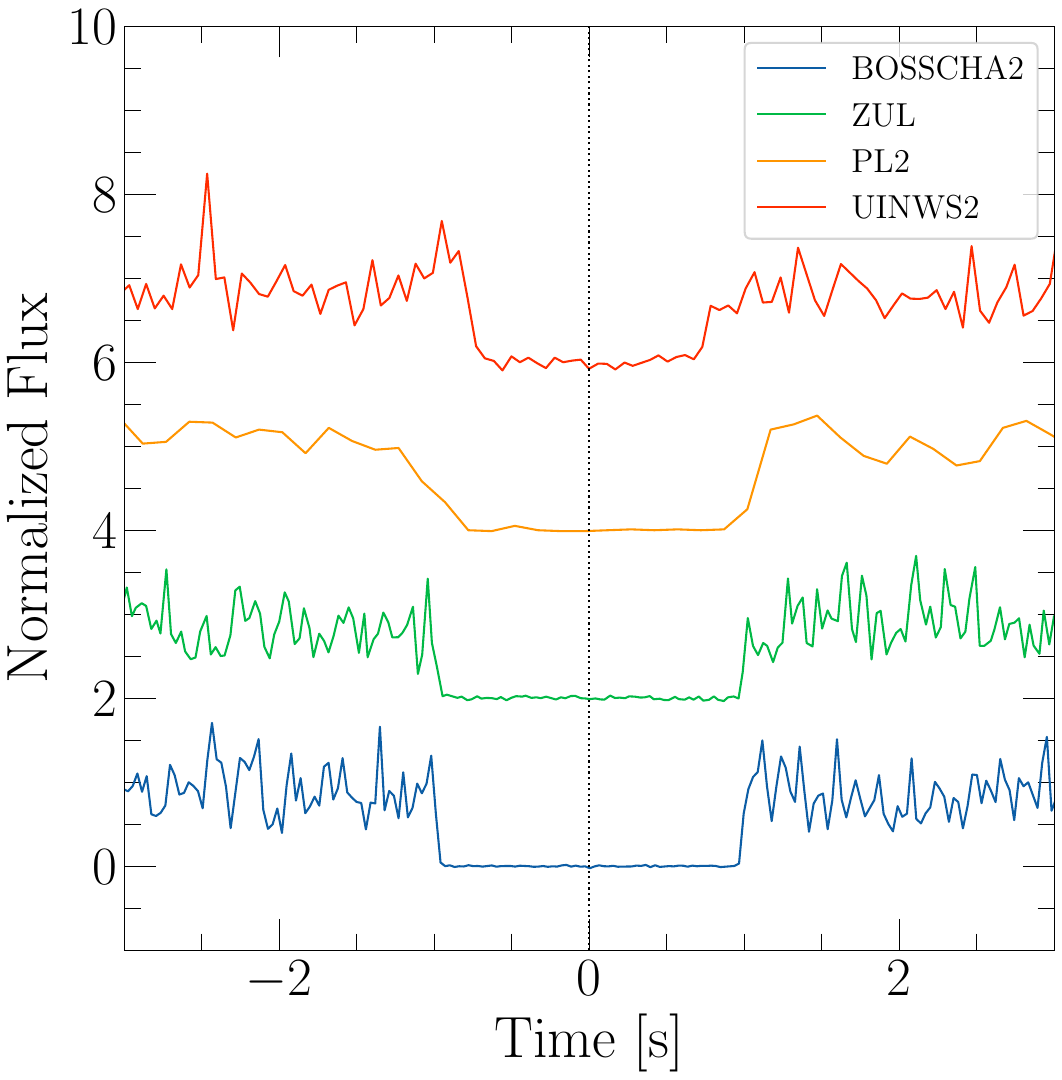}
	\caption{Time-series photometry for the four successful chords. The light curves are temporally aligned to their respective mid-event times and displaced along the vertical axis for visual clarity.}\label{fig:positive_lcs} 
\end{figure}

For every recorded frame, a set of synthetic apertures spanning a range of radii was trialed. The radius that minimized the dispersion of the out-of-occultation baseline was retained. The baseline of each normalized light curve was then fitted with a fifth-order Legendre polynomial. This removes residual transparency and airmass trends over the recording. The fit was performed only on the data outside the occultation window, without absorbing the ingress and egress or biasing the contact times \citep{Gomes2015}. Per-point photometric uncertainties were estimated directly from the scatter of this baseline. These provide the $\sigma_i$ used in the light-curve modelling of Section~\ref{sec:data}. The light curves for the four positive chords, aligned to their respective mid-event times, are shown in Fig.~\ref{fig:positive_lcs}.

\section{Data Reduction and Timing Analysis}
\label{sec:data}

\subsection{Polychromatic Forward Modeling of the Occultation Light Curve}

The extraction of precise physical dimensions from stellar occultation light curves requires a forward model that accurately reproduces the complex diffraction geometry of the event \citep[e.g.,][]{roques1987}. We modeled the observed normalized flux by generating synthetic light curves that account for Fresnel diffraction, the finite angular size of the background star, and instrumental smearing effects.

For an occulting body (or circum-asteroidal feature) of physical width $W$ and maximum opacity $A$, the baseline monochromatic diffraction pattern is governed by the Fresnel scale, $F_\lambda = \sqrt{\lambda \rho / 2}$, where $\lambda$ is the observing wavelength and $\rho$ is the topocentric distance to the object derived from JPL Horizons ephemerides, queried via the \texttt{astroquery.jplhorizons} module (\citealp{Ginsburg2019}). However, because the spatial frequency of Fresnel fringes is strictly wavelength-dependent ($F_\lambda \propto \sqrt{\lambda}$), integrating across a broad bandpass causes higher-order fringes to destructively interfere, a phenomenon known as chromatic smearing \citep{widemann2009}. Relying on a single ``effective wavelength'' can therefore artificially preserve high-order ringing and bias the derived feature widths.

To rigorously model this, we employed a polychromatic approach. We defined a normalized total spectral response function, $S(\lambda) = \mathrm{SED}(\lambda) \times T(\lambda) \times \mathrm{QE}(\lambda)$, constructed from the product of the background star’s spectral energy distribution, the filter transmission curve, and the wavelength-dependent quantum efficiency of the detector. Monochromatic point-source patterns were computed over a discretized wavelength grid spanning the bandpass and numerically integrated according to the weights of $S(\lambda)$.

Because the background star is not a true point source, its finite angular diameter further dampens the diffraction fringes. The polychromatic point-source pattern was convolved with the 1D projection of a uniform circular stellar disk of diameter $D_*$. Finally, to model temporal smearing, the spatial profile was convolved with a boxcar kernel corresponding to the camera’s finite exposure time $\Delta t_{\mathrm{exp}}$, scaled by the apparent shadow velocity $v$ across the observer's plane.

\subsection{Parameter Estimation and Bayesian Inference}

Occultation light curves are characterized by severe non-linearities and parameter degeneracies, most notably between the stellar diameter, the impact parameter, and the feature width. To systematically map this complex parameter space, we implemented a two-stage optimization and Bayesian inference framework.

Initial parameter estimates were obtained using a Differential Evolution (DE) global optimization algorithm \citep{Storn1997}. This heuristic approach is highly resistant to local likelihood maxima, ensuring the solver identifies the global minimum basin. These optimized parameters subsequently seeded a Trust Region Reflective (TRF) least-squares routine to rapidly converge on the exact maximum a posteriori (MAP) estimate.

To derive robust uncertainties and explore parameter covariances, the MAP parameters initialized an affine-invariant Markov Chain Monte Carlo (MCMC) ensemble sampler \citep{goodman2010}, implemented via the \texttt{emcee} Python package \citep{ForemanMackey2013}. We assumed Gaussian, uncorrelated photometric uncertainties, defining the unnormalized log-posterior probability as proportional to the log-likelihood:
\begin{equation}
	\ln \mathcal{L}(\bm{\theta}) = -\frac{1}{2} \sum_{i} \left( \frac{f_i - M_i(\bm{\theta})}{\sigma_i} \right)^2
\end{equation}
where $f_i$ is the observed normalized flux, $\sigma_i$ is the photometric uncertainty, and $M_i(\bm{\theta})$ is the forward model evaluated at time $t_i$ for the parameter vector $\bm{\theta}$. We adopted uniform, physically bounded priors (e.g., $D_* > 0$, $A \in [0,1]$) to restrict the sampler to physically permissible parameter spaces.

\subsection{Derivation of Physical Quantities}

The final best-fit parameter values and their corresponding $1\sigma$ uncertainties were drawn from the 50th, 16th, and 84th percentiles of the marginalized posterior distributions, respectively.

To account for the possibility of unmodeled background fluctuations or non-opaque limb features, $A$ was left as a free parameter in our forward model. For features where $W$ is comparable to or narrower than the Fresnel scale and the projected stellar diameter, $W$ and $A$ become highly degenerate. While characterizing the physical opacity of the body is outside the scope of this astrometric analysis, computing $A$ alongside the kinematic parameters directly within the MCMC framework naturally propagates the covariant errors between the feature's width, opacity, and timing. Consequently, the final reported astrometric uncertainties are robustly marginalized over this highly degenerate parameter space.

\begin{figure*}[ht]
	\centering
	\begin{subfigure}[]{0.45\textwidth}
		\centering
		\includegraphics[height=0.18\textheight]{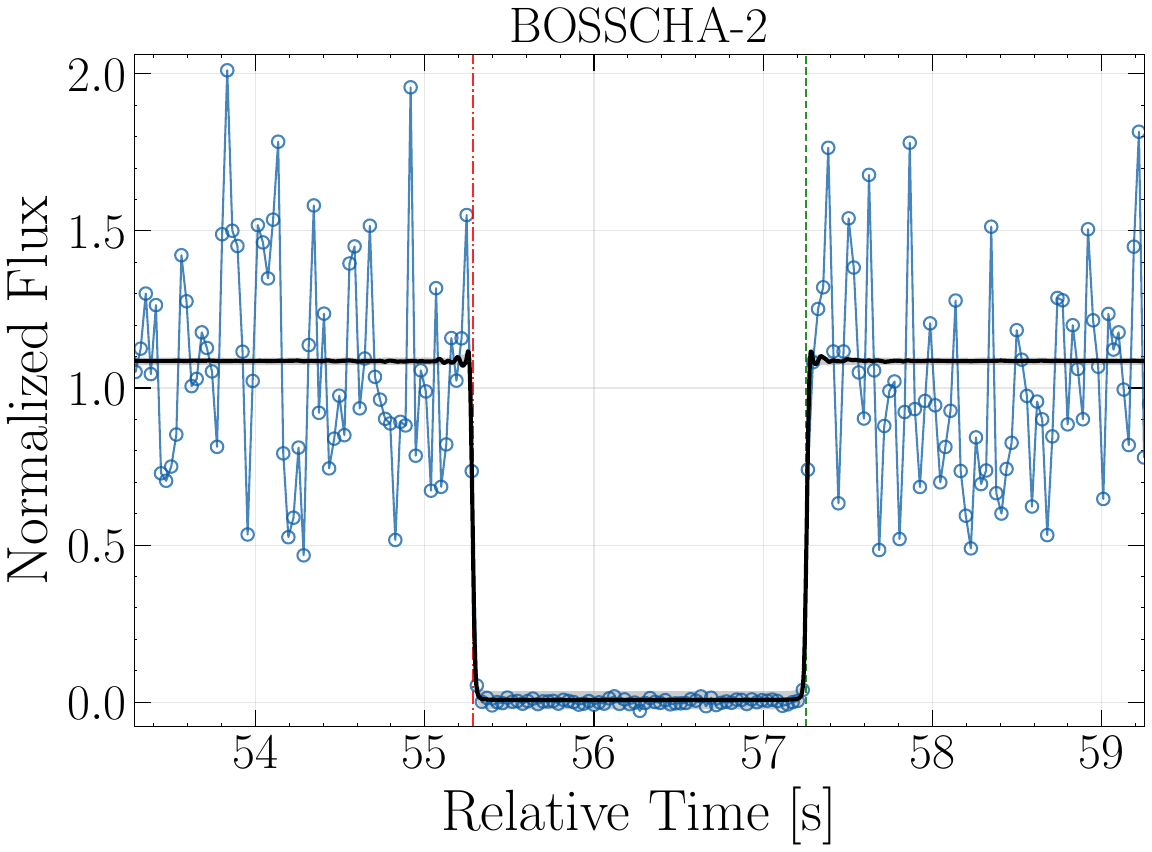}
	\end{subfigure}\hfil
	\begin{subfigure}[]{0.45\textwidth}
		\centering
		\includegraphics[height=0.18\textheight]{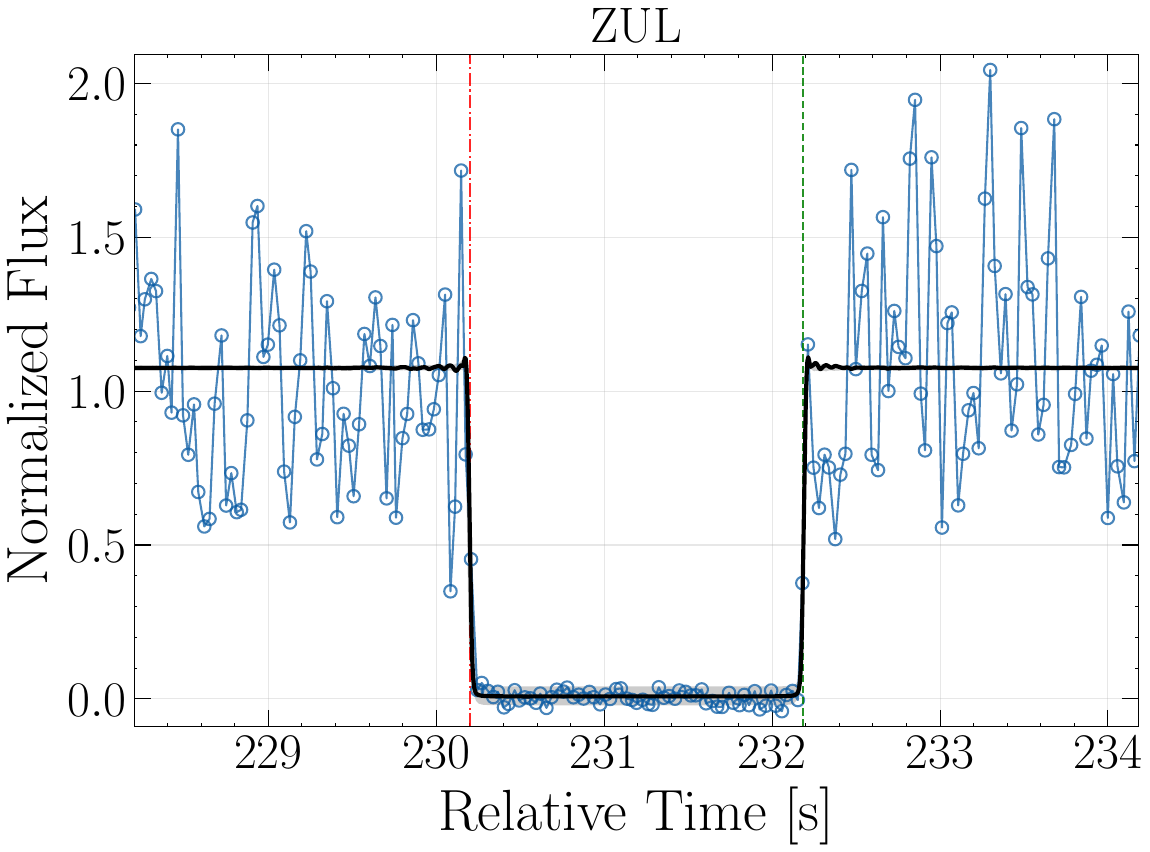}
	\end{subfigure}
	
	\vspace{1em}
	
	\begin{subfigure}[]{0.45\textwidth}
		\centering
		\includegraphics[height=0.18\textheight]{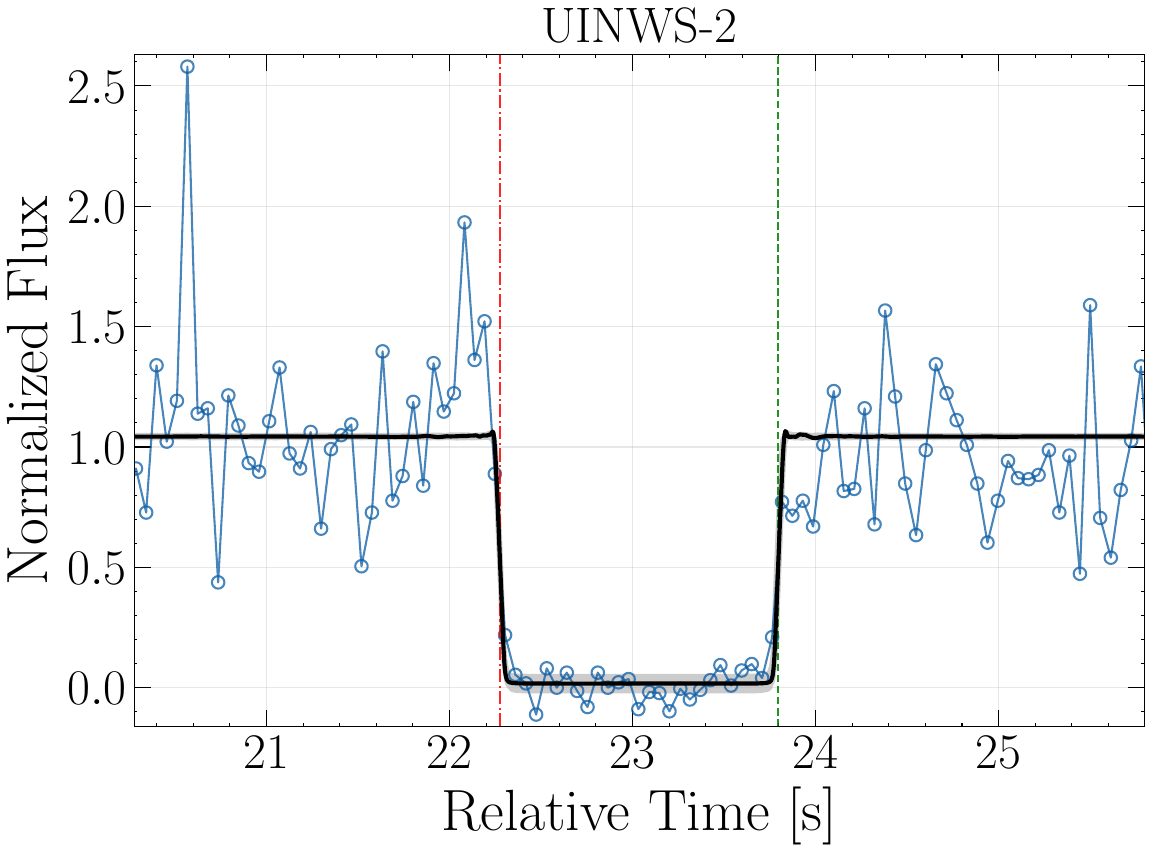}
	\end{subfigure}\hfil
	\begin{subfigure}[]{0.45\textwidth}
		\centering
		\includegraphics[height=0.18\textheight]{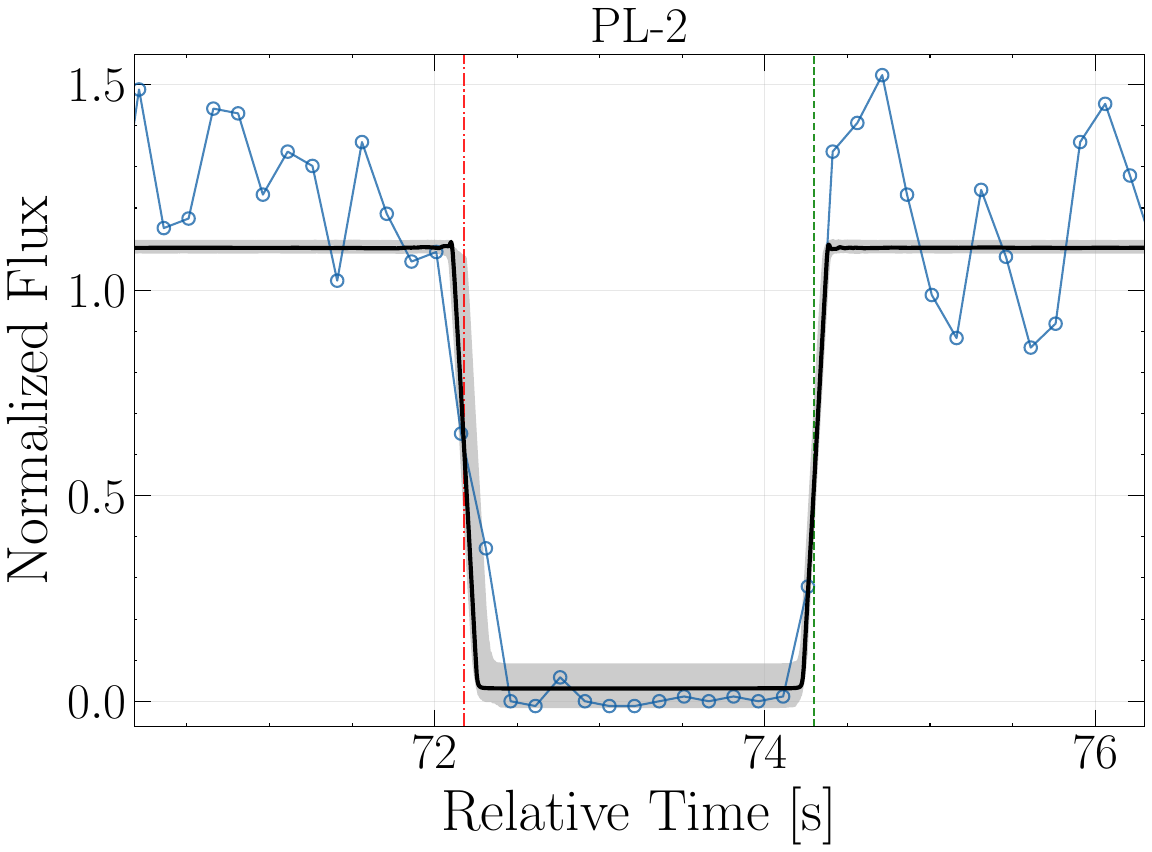}
	\end{subfigure}
	\caption{Best-fit models for the observed light curves of all positive detections. The thick black lines represent the Bayesian median fits, while the open blue circles indicate the observed data points. The vertical red dash-dotted and green dashed lines indicate the immersion and emersion times, respectively. The gray shaded regions represent the 68\% credible intervals of the fits.}\label{fig:lc_fit_plots} 
\end{figure*}

Our Bayesian framework successfully converged for all positive detections, yielding precise ingress and egress timings for each station (Table~\ref{tab:lc_fit_parameters}), with the best-fitting models are displayed in Fig.~\ref{fig:lc_fit_plots}.

\begin{deluxetable*}{lcccc}
	\tablecaption{Best-fit parameters derived from the light-curve analysis (see Fig.~\ref{fig:lc_fit_plots}).\label{tab:lc_fit_parameters}}
	\tablewidth{0pt}
	\tablehead{
		\colhead{Station} & \colhead{Immersion time} & \colhead{Emersion time} & \colhead{Chord duration} & \colhead{Chord length} \\
		\colhead{} & \colhead{(UTC)} & \colhead{(UTC)} & \colhead{(s)} & \colhead{(km)}
	}
	\startdata
	BOSSCHA-2 & 12:42:33.258$^{+0.008}_{-0.008}$ & 12:42:35.228$^{+0.008}_{-0.009}$ & $1.970^{+0.011}_{-0.012}$ & 47.85$^{+0.28}_{-0.30}$ \\
	ZUL & 12:42:33.266$^{+0.012}_{-0.018}$ & 12:42:35.252$^{+0.010}_{-0.008}$ & $1.986^{+0.020}_{-0.014}$ & 48.26$^{+0.44}_{-0.38}$ \\
	PL-2 & 12:41:57.516$^{+0.083}_{-0.030}$ & 12:41:59.635$^{+0.024}_{-0.024}$ & $2.119^{+0.039}_{-0.087}$ & 51.22$^{+1.07}_{-1.87}$ \\
	UINWS-2 & 12:41:57.414$^{+0.018}_{-0.018}$ & 12:41:58.934$^{+0.017}_{-0.023}$ & $1.520^{+0.025}_{-0.029}$ & 36.77$^{+0.66}_{-0.65}$ \\
	\enddata
\end{deluxetable*}


\section{Chord Fitting and Silhouette Reconstruction}
\label{sec:chord}

To determine the physical profile of the occulting body, topocentric timing observations were projected onto the Besselian fundamental plane i.e. a geocentric Cartesian coordinate system defined such that its $xy$-plane is strictly orthogonal to the line of sight toward the target star. The reduction pipeline follows a rigorous astrometric transformation sequence to project both the moving observer and the asteroid's shadow onto this common 2-dimensional reference frame \citep{Herald2020}.

\subsection{Stellar and Ephemeris Preparation}

The coordinates of the target star were sourced from the \textit{Gaia} Data Release 3 \citep[DR3;][]{GaiaDR3_2023} catalog. The reference astrometry (Right Ascension, Declination, parallax, and proper motions) was propagated from the catalog epoch (J2016.0) to the epoch of the occultation event, strictly maintaining the International Celestial Reference System (ICRS) framework. Simultaneously, the ephemerides of the asteroid were retrieved from the JPL Horizons system through the \texttt{astroquery.jplhorizons} module for a time window bracketing the event. To ensure temporal consistency, the queried ephemerides utilized astrometric positions, inherently accounting for planetary aberration and the light-travel time from the asteroid to the Earth.

\subsection{Observer Coordinate Transformation}

The topocentric coordinates of each observing station, initially defined by geodetic longitude ($\lambda$), latitude ($\phi$), and elevation relative to the WGS84 reference ellipsoid, were converted into geocentric Cartesian coordinates. For every recorded immersion ($t_\text{in}$) and emersion ($t_\text{eg}$) time, the observer's instantaneous geocentric position was projected onto the fundamental plane \citep{Urban2013}. 

This yields the observer coordinates $(\xi, \eta)$, calculated as a function of the Earth's rotation and the local hour angle ($H$) of the target star at the recorded contact times:
\begin{align}
	\xi &= \Delta \cos\phi' \sin H \\
	\eta &= \Delta \sin\phi' \cos\delta - \Delta \cos\phi' \sin\delta \cos H
\end{align}
where $\Delta$ is the geocentric distance, $\phi'$ is the geocentric latitude, and $\delta$ is the declination of the star.

\subsection{Kinematic Reduction and Profile Reconstruction}

To reconstruct the asteroid's profile, the timing data must be transformed into an asteroid-fixed reference frame. The position of the asteroid's shadow center on the fundamental plane, $(x_\text{c}, y_\text{c})$, was derived continuously from the interpolated JPL Horizons ephemerides. For each observing chord, the physical position of the asteroid's limb relative to its center of mass was determined by calculating the difference between the observer's projected position and the shadow's predicted center at the exact moments of immersion and emersion:
\begin{align}
	f(t) &= \xi(t) - x(t) \\
	g(t) &= \eta(t) - y(t)
\end{align}
The resulting $(f, g)$ coordinates for $t_\text{in}$ and $t_\text{eg}$ define the endpoints of each observed chord across the asteroid's profile.

\subsection{Fundamental Plane Parametrization and Constraints}

The pipeline models the projected two-dimensional silhouette of the occulting body on the fundamental plane as an ellipse. This assumption models the shadow cast by a triaxial ellipsoid under distant stellar illumination. The geometry is defined by five parameters: the center coordinates of the shadow $(x_c, y_c)$, the semi-major and semi-minor axes $(a, b)$, and the position angle of the major axis $\theta$.

To ensure physically viable geometries and prevent the sampler from exploring degenerate, needle-like topologies, strict hard boundaries are enforced on the oblateness. The axis ratio is constrained to $0.1 \le b/a \le 1.0$. The observer trajectories are mapped onto this shadow-centric reference frame, categorizing the spatial data into positive detections (ingress/egress contact points) and negative detections (continuous path segments defining observational misses).

\subsection{Positive Chord Likelihood via First-Order Algebraic Distance}

Calculating the exact geometric orthogonal distance from an observed datum to a trial ellipse requires finding the roots of a quartic polynomial, which is computationally prohibitive within a highly iterative MCMC architecture. To resolve this, the fitting engine employs a first-order algebraic distance approximation, mathematically equivalent to the Sampson error or Taubin's method \citep{Taubin1991}.

For a given trial ellipse, the coordinate of each contact point $(x_i, y_i)$ is rotated by $-\theta$ and normalized by the semi-axes $(a, b)$ into a unit-circle frame $(X, Y)$. The algebraic distance $f$ and its spatial gradient magnitude $\|\nabla f\|$ are evaluated to linearly approximate the true geometric distance. The likelihood function for the $N$ positive detections is then formulated as a weighted $\chi^2$ statistic:
\begin{equation}
	\chi^2_{\text{pos}} = \sum_{i=1}^{N} w_i \left( \frac{f_i}{\|\nabla f_i\| \sigma} \right)^2
\end{equation}
where $\sigma$ is the global positional uncertainty parameter. The inclusion of the statistical weight array ($w_i$) is mathematically critical; it allows the model to naturally synthesize heterogeneous multi-chord data, properly weighting high-fidelity observations (e.g., hardware-level GPS timestamping with minimal deadtime) against standard NTP synchronized stations.

\subsection{Negative Chord Topography and Continuous Penalty Functions}
Negative observations provide upper limits on the target's physical dimensions. Rather than discarding these constraints or treating them as discrete points, the algorithm rigorously models them as finite line segments representing the observer's continuous trajectory across the fundamental plane.

During each evaluation, the negative line segment is transformed via an affine mapping into the normalized $(U, V)$ unit-circle frame of the trial ellipse. The algorithm computes the minimum squared distance $d^2$ from the segment to the origin. A condition of $d^2 \le 1$ indicates a physical impossibility: the proposed finite shadow path has enveloped an observer who reported a negative detection.

To enforce this spatial exclusion, a discontinuous infinite penalty ($-\infty$) is avoided, as step functions can trap affine-invariant samplers in topological dead zones. Instead, the algorithm applies a continuous, steep half-quadratic penalty function proportional to the penetration depth:
\begin{equation}
	\Delta\chi^2_{\text{pen}} = \sum_{j} 10^8 \times (1 - d_j)^2 \quad \text{for} \quad d_j \le 1
\end{equation}
This pseudo-$\chi^2$ penalty strictly enforces the bounding limits of the negative chords while maintaining a mathematically differentiable posterior landscape, aiding the global optimizer in converging toward valid geometries. 

\subsection{Bayesian Inference and Global Optimization}

The inverse problem is resolved within a Bayesian framework to rigorously quantify parameter uncertainties, correlations, and marginal distributions. The log-posterior probability $\ln \mathcal{L}_{\text{post}}$ is constructed from the data likelihood, the continuous negative penalties, the normalization constant of the variance, and an explicit shape prior:
\begin{equation}
	\ln \mathcal{L}_{\text{post}} = -\frac{1}{2} \left( \chi^2_{\text{pos}} + \Delta\chi^2_{\text{pen}} + N \ln(2\pi\sigma^2) \right) + \ln P_{\text{prior}}
\end{equation}
To address structural degeneracies inherent in elliptical fitting, particularly when chords are sparse or non-uniformly distributed across the object's hemisphere, a soft Gaussian prior is applied favoring circularity:
\begin{equation}
	\ln P_{\text{prior}} = -0.5 \left( \frac{b/a - 1.0}{0.2} \right)^2
\end{equation}
This prior reflects the physical expectation that the projected silhouettes of smaller, unconstrained solar system bodies do not generally exhibit extreme elongations unless strongly supported by the chord distribution.

\begin{figure}[ht]
	\centering
	\includegraphics[width=0.45\textwidth]{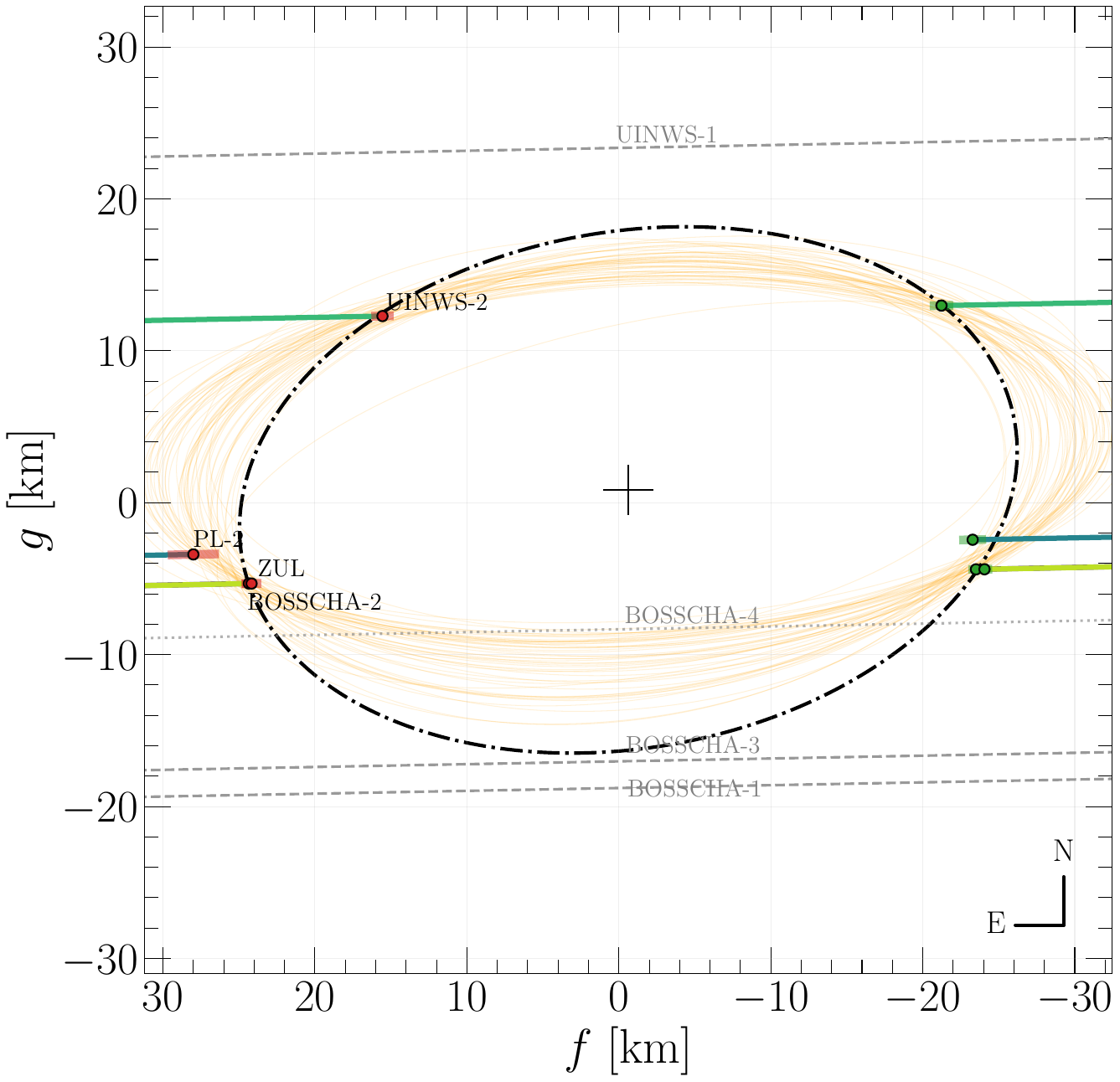}
	\caption{Best-fitting limb model overlaid on the occultation chords. The black dash-dotted curve denotes the best-fitting limb. Thick colored segments represent the positive occultation chords recorded by the individual observers listed in Table~\ref{tab:lc_fit_parameters}, whereas dashed gray lines indicate negative chords. Dotted gray lines correspond to observation attempts for which no data were available. Red and green segments mark the ingress and egress timing uncertainties at the chord extremities, respectively. Thin orange curves delineate the 68\% credible region of the reconstructed limb. The BOSSCHA-2 and ZUL stations are separated a few meters away, making them appear superimposed in the figure.}\label{fig:multichords_fit}
\end{figure}

The parameter space was explored using a two-stage hybrid optimization framework. First, a DE algorithm was employed to efficiently navigate the highly multimodal parameter space and identify the global MAP solution. This best-fit solution was subsequently used to initialize an affine-invariant ensemble MCMC sampler for posterior sampling. To ensure robust sampling, a strict initialization criterion was applied so that all walkers began within valid, finite regions of the posterior probability density, thereby promoting stable burn-in behavior and reliable convergence of the production chains. The best-fitting model of the occultation chords and the corresponding parameter estimates are presented in Fig.~\ref{fig:multichords_fit} and Table~\ref{tab:strenua_parameters}, respectively.

\begin{deluxetable*}{llc}
	\tablecaption{Summary of Astrometric and Physical Parameters for (1201) Strenua\label{tab:strenua_parameters}}
	\tablewidth{0pt}
	\tablehead{
		\colhead{Parameter} & \colhead{Value} & \colhead{Unit}
	}
	\startdata
	\multicolumn{3}{l}{\textit{\textbf{Reference Information}}} \\
	Observation Epoch (Mid-time) & 2026-04-26 12:42:16.313 & UTC \\
	\hline
	\multicolumn{3}{l}{\textit{\textbf{Derived Astrometry}}} \\
	Asteroid Right Ascension ($\alpha_\text{J2000.0}$) & $07^\text{h}24^\text{m}27.64797^\text{s}$ & hours \\
	Asteroid Declination ($\delta_\text{J2000.0}$) & $+15^\circ31'01.9123''$ & deg \\
	Fundamental Plane Offset ($\xi$, $\eta$) & $-23.06, +8.33$ & km \\
	Right Ascension Uncertainty ($\sigma_{\alpha^*}$)\tablenotemark{a} & $1.14$ & mas \\
	Declination Uncertainty ($\sigma_{\delta}$) & $0.55$ & mas \\
	Astrometric Correlation ($\rho_{\alpha* \delta}$) & $-0.203$ & -- \\
	\hline
	\multicolumn{3}{l}{\textit{\textbf{Physical Profile (Posterior Fit)}}\tablenotemark{b}} \\
	Area-Equivalent Diameter ($D_\text{eq}$) & $39.86^{+2.06}_{-1.66}$ & km \\
	Apparent Semi-major Axis ($a'$) & $27.49^{+3.90}_{-1.65}$ & km \\
	Apparent Semi-minor Axis ($b'$) & $14.33^{+2.37}_{-2.69}$ & km \\
	Apparent Oblateness ($\epsilon'$) & $0.48^{+0.15}_{-0.12}$ & -- \\
	Position Angle ($\theta$) & $95.3^{+4.9}_{-2.7}$ & deg (E of N) \\
	\enddata
	\tablenotetext{a}{The right ascension uncertainty contains the cosine of the declination correction: $\alpha^* \equiv \Delta\alpha \cos\delta$.}
	\tablenotetext{b}{Physical profile uncertainties represent the median and $1\sigma$ bounds derived from the MCMC posterior distribution.}
\end{deluxetable*}

\subsection{Comparison with previous diameter determinations}

The area-equivalent diameter derived from the multi-chord occultation ($D_\text{eq} = 39.86^{+2.06}_{-1.66} \text{ km}$, cf. Table~\ref{tab:strenua_parameters}) provides a direct, model-independent measurement that contextualizes and refines prior independent size estimates for (1201) Strenua. Before this work, the size of this asteroid was inferred almost exclusively from thermal infrared radiometric observations or from single-chord occultation constraints, rather than from direct multi-chord measurements of the projected limb.

The earliest widely used diameter originates from the IRAS Minor Planet Survey \citep{Tedesco2002}, which derived a diameter of $34.86 \pm 3.50$~km using the Standard Thermal Model (STM; \citealp{Lebofsky1986}). Later, the WISE/NEOWISE survey \citep{Masiero2011} independently obtained a diameter of $35.825 \pm 0.240$~km using the Near-Earth Asteroid Thermal Model (NEATM; \citealp{Harris1998}). Although the IRAS and WISE analyses adopt different thermal models and benefit from different infrared wavelength coverage, both converged on essentially the same physical size, with their difference of $\sim 1$~km falling well within the $10\%$ uncertainty of the IRAS determination and the typical systematic limits of thermal radiometry.

An additional independent constraint is provided by a single-chord occultation observed in 2019 by K.~Getrost (2026, private communication; Occult4 database). Although insufficient for reconstructing the asteroid's silhouette, that event yielded a chord length of $37.2$~km. Single-chord events cannot uniquely determine the projected figure because the observed chord may pass anywhere between the center and the limb of the asteroid. Consequently, the measurement acts as a strict lower bound on the maximum projected dimension along that chord.

Interestingly, while our multi-chord equivalent diameter is approximately $11\%$ larger than the WISE NEATM determination, Kai Getrost's single-chord lower limit independently confirms that Strenua's maximum projected dimension must exceed the infrared spherical fits. This discrepancy is readily explained by our resolved projected geometry. The multi-chord fit reveals that Strenua is highly elongated, with an apparent oblateness derived directly from the posterior distribution to be $\epsilon' = 0.48^{+0.15}_{-0.12}$ (cf. Table~\ref{tab:strenua_parameters}; corresponding to an axis ratio of $a'/b' \approx 1.9$). Because of this severe departure from a spherical shape, standard thermal models will derive varying effective diameters depending on the rotational phase and projected cross-section presented at the epoch of the infrared observations. Occultation limb-mapping, by contrast, directly measures the instantaneous spatial extent without relying on assumed thermal emissivity or beaming parameters. The $\sim 4$~km difference between the radiometric sizes and our multi-chord silhouette is therefore entirely consistent with the highly triaxial, collisionally evolved nature of Strenua.

\subsection{Astrometric Solution and Uncertainty Analysis}

The asteroid's astrometric position was obtained by applying the best-fitting offset of the reconstructed occultation shadow center to the propagated Gaia DR3 position of the occulted star. The asteroid's center of mass was found to be offset from the stellar position by $\xi = -23.06$~km and $\eta = 8.33$~km (cf. Table~\ref{tab:strenua_parameters}). Converting these physical offsets into angular displacements and applying them yields an absolute astrometric position for the asteroid in the International Celestial Reference System (ICRS / J2000.0) of $\alpha = 07^\text{h}24^\text{m}27.64797^\text{s}$ and $\delta = +15^\circ31'01.9123''$. To preserve both the variances of the individual coordinates and their statistical correlation, the uncertainty of this derived position is parameterized using the full covariance matrix rather than independent, one-dimensional uncertainties. This approach provides a complete description of the two-dimensional astrometric uncertainty associated with the occultation geometry. The covariance matrix of the derived asteroid position, defined in terms of $\alpha^* \equiv \Delta\alpha \cos\delta$ and the declination offset $\delta$, is
\begin{equation}
	C_{\alpha^*\delta} = \begin{pmatrix} 1.29 & -0.12 \\ -0.12 & 0.30 \end{pmatrix} \, \text{mas}^2
\end{equation}
where the diagonal elements correspond to the variances of the right ascension and declination coordinates, and the off-diagonal elements represent their covariance. The corresponding marginal uncertainties, obtained from the square roots of the diagonal elements, are $1.14$~mas in right ascension and $0.55$~mas in declination. These values demonstrate that the multichord occultation constrains the asteroid position with milliarcsecond-level precision in both coordinates.

The covariance term corresponds to a correlation coefficient of $\rho_{\alpha* \delta} = -0.203$, indicating a weak negative correlation between the right ascension and declination estimates. Such correlations naturally arise because both coordinates are estimated simultaneously from the same shadow geometry and are therefore not statistically independent. However, the relatively small magnitude of the correlation indicates that the two coordinates are only weakly coupled.

The covariance matrix is symmetric and positive definite, defining a physically meaningful two-dimensional uncertainty ellipse. Eigenvalue decomposition yields principal-axis uncertainties of $1.14$~mas and $0.53$~mas, with the major axis rotated by approximately $7^\circ$ with respect to the right ascension axis. This small rotation angle, combined with the weak off-diagonal correlation, demonstrates that the principal axes of the uncertainty ellipse remain closely aligned with the equatorial coordinate system. However, the uncertainty in right ascension is approximately a factor of two larger than in declination, resulting in an error ellipse that is distinctly elongated in the East-West direction.

By representing the astrometric uncertainty via its full covariance matrix, all second-order statistical information contained in the posterior distribution is preserved. This rigorous representation ensures that the statistical coupling between the coordinate estimates is correctly accounted for when the astrometry is propagated into subsequent analyses, such as ephemeris refinement, orbit determination, or combination with independent measurements.


\section{Geometric Albedo and Compositional Classification}
\label{sec:albedo}

The geometric albedo $p_\mathrm{V}$ of an airless body is a critical diagnostic metric for evaluating surface composition, texture, and porosity. It serves as a primary taxonomic classifier distinguishing primitive, low-albedo carbonaceous bodies from evolved, higher-albedo silicaceous or metallic ones, and is essential for modeling regolith properties that drive non-gravitational perturbations like the Yarkovsky and YORP effects. Furthermore, $p_\mathrm{V}$ is indispensable for determining asteroid sizes; given an independent diameter $D$ estimate from thermal radiometry or stellar occultations \citep{Cellino2015, Mottola2023}, $p_\mathrm{V}$ is mathematically coupled to the Johnson $V$-band absolute magnitude, $H_\mathrm{V}$, via the standard scaling relationship:
\begin{equation}\label{eq:albedo_absmag}
	\log_{10}(D) = 3.1236 - 0.2H_\mathrm{V} - 0.5\log_{10}(p_\mathrm{V})
\end{equation}

\subsection{Phase Curve Parameters}\label{subsec:phasecurve_params}
Because observing at exactly zero phase angle is impossible from Earth, the absolute magnitude $H$ must be extrapolated by fitting a phase function model to observations across various phase angles \citep{Wilawer2024}. To resolve the failure of the legacy $H, G$ system \citep{Bowell1989} to accurately capture the non-linear opposition surge, the IAU adopted the three-parameter $H, G_1, G_2$ system \citep{Muinonen2010} in 2012 to isolate the linear phase slope from the opposition effect. The morphology of this phase curve is strongly governed by regolith light-scattering physics: higher-albedo objects exhibit flatter curves driven by multiple scattering, whereas low-albedo objects show steeper curves dominated by single-particle scattering and shadow-hiding. Consequently, the $G_1$ and $G_2$ parameters, which dictate the function's slope and curvature, naturally cluster asteroids into taxonomic domains and serve as reliable empirical proxies for $p_\mathrm{V}$, enabling robust albedo estimation purely from well-sampled phase curves when independent thermal or polarimetric data are unavailable.

In this work, we utilize ATLAS data \citep{Tonry2018} in the cyan ($c$) and orange ($o$) filters, retrieved from the Minor Planet Center (MPC). To accurately weight the photometric data in our analysis, per-observation uncertainties were incorporated by retrieving records from the MPC in the Astrometry Data Exchange Standard (ADES; \citealp{Chesley2017}). Because legacy 80-column MPC observation tables do not natively report photometric errors, querying the modern ADES-formatted Observations API is necessary to extract the 1-sigma root-mean-square magnitude (\texttt{rmsMag}) associated with each observational epoch. To ensure robust automated ingestion, our reduction pipeline sanitizes the API’s JSON data frames to account for known schema inconsistencies. Specifically, the pipeline normalizes variable key casing and strips non-standard UTC timezone characters (trailing "Z" indicators) from the ISO-8601 temporal strings. These sanitized epochs are subsequently parsed into Julian Dates using the \texttt{astropy.time} package \citep{TheAstropyCollaboration2022}. This automated cleaning allows the exact reported uncertainties to be confidently cross-matched with the historical observation tables, ensuring we only rely on coarse, survey-specific nominal priors when authentic ADES measurements are genuinely missing from the archival record.

For each data point, we query the JPL Horizons system via the \texttt{astroquery.jplhorizons} module. By providing the specific epoch and observatory location, we extract the precise heliocentric ($r$) and topocentric ($\rho$) distances. 

To account for rotational brightness variations, the $c$- and $o$-band photometric measurements were modeled simultaneously as a function of solar phase angle and rotational phase. To constrain the simultaneous fit, a common synodic rotation period $P$ was shared across both passbands. Assuming a roughly constant aspect angle over the observation epoch, the rotational modulation for each passband was represented by an independent $k$-order Fourier series,
\begin{equation}
	V_{\mathrm{rot}}(\phi) 	= 	\sum_{i=1}^{k} 	\left[ 	A_i \cos(2\pi i\phi) 	+ 	B_i \sin(2\pi i\phi) 	\right], 	\label{eq:rot_mod}
\end{equation}
where the rotational phase is defined as $\phi = (t - t_0) / P \pmod 1$, given the light-travel-time-corrected observation epoch $t$ and a zero-point reference epoch $t_0$. The band-dependent harmonic coefficients $A_i$ and $B_i$ describe the shape and amplitude of the rotational light curve. The rotational component was incorporated directly into the $H,G_1,G_2$ phase-function model such that the predicted reduced magnitude is given by
\begin{equation}
	m(\alpha,\phi) 	= 	m_{\mathrm{HG_1G_2}}(\alpha) 	+ 	V_{\mathrm{rot}}(\phi) 	\left(1+c_{\mathrm{amp}, b}\alpha\right), 	\label{eq:reduced_mag}	
\end{equation}
where $m_{\mathrm{HG_1G_2}}(\alpha)$ is the $H,G_1,G_2$ phase-function prediction, $\alpha$ is the solar phase angle (in degrees), and $c_{\mathrm{amp}, b}$ is a band-specific free parameter describing the linear dependence of the rotational amplitude on phase angle. While the rotation period $P$ was constrained globally, the amplitude-scaling coefficient $c_{\mathrm{amp}, b}$ and the Fourier coefficients were fitted independently for each passband. This formulation successfully accounts for the empirically observed Phase Angle-Amplitude (PAA) effect, where rotational light-curve amplitude increases with increasing phase angle, a behavior commonly attributed to enhanced shadowing and surface roughness.

\begin{figure*}[ht]
	\centering
	\begin{subfigure}[]{0.45\textwidth}
		\centering
		\includegraphics[height=0.22\textheight]{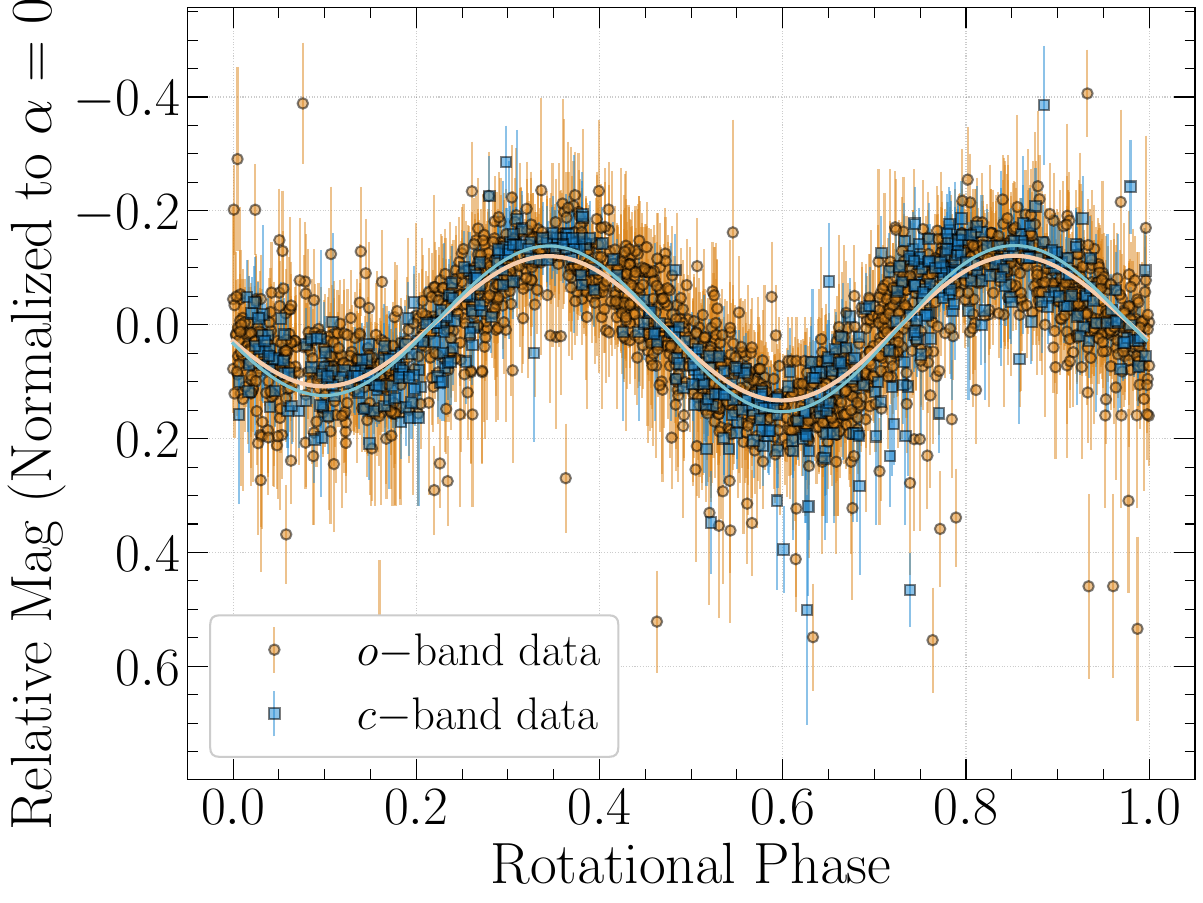}
	\end{subfigure}\hfil
	\begin{subfigure}[]{0.45\textwidth}
		\centering
		\includegraphics[height=0.22\textheight]{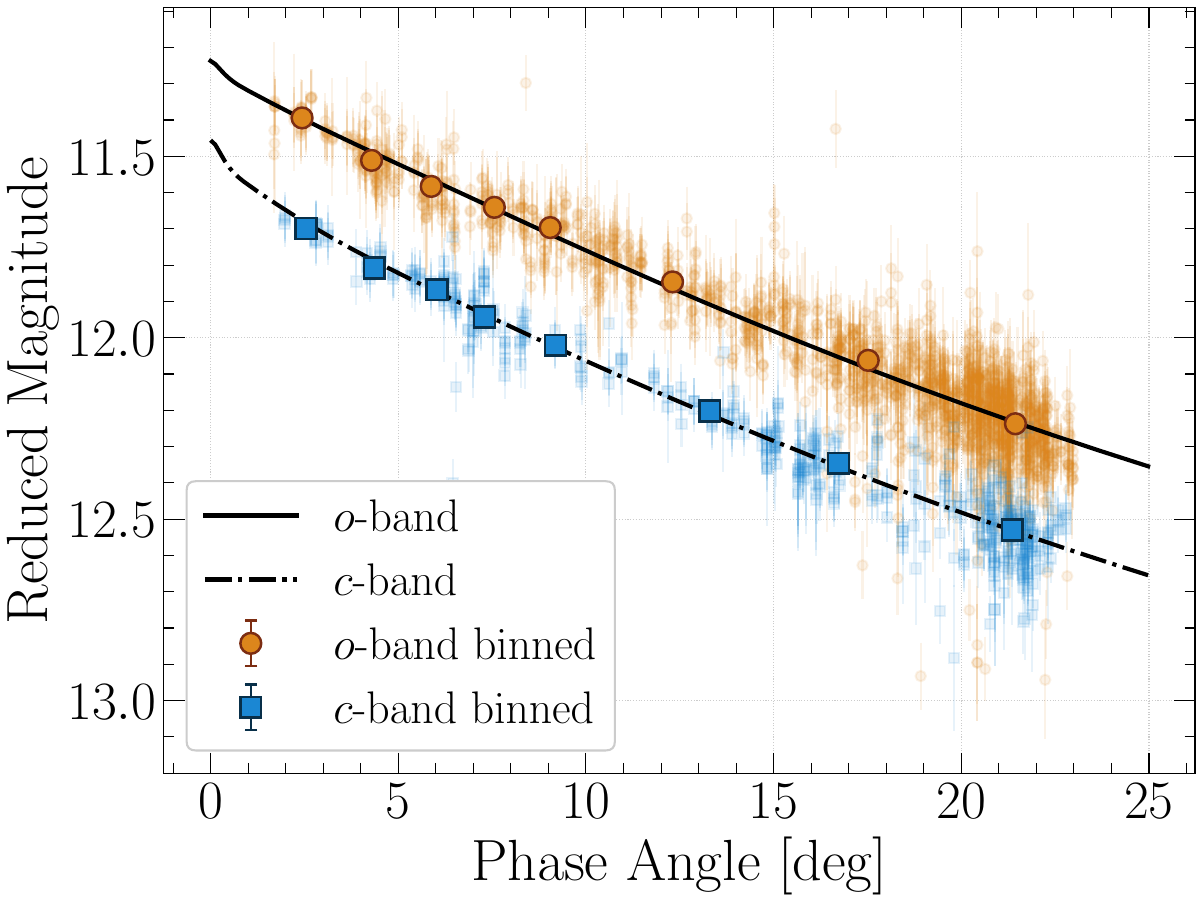}
	\end{subfigure}
	\caption{Left: ATLAS $c$- and $o$-band photometric data for (1201) Strenua, phased to the best-fit rotation period of $14.77495^{+0.00001}_{-0.00002}$ h. Right: Corresponding phase-curve fits obtained using the $H,G_1,G_2$ photometric system for each band.}\label{fig:multiband_diag} 
\end{figure*}

The optimization of the joint parameter space, encompassing both the rotational model and the $H, G_1, G_2$ phase function, proceeded in two sequential stages. To identify a high-likelihood region, we first initialized the fitting routine using a DE global optimization algorithm. This global solution subsequently seeded an \texttt{emcee} MCMC analysis to simultaneously infer the posterior distributions of all model parameters. Within this joint framework, the shared rotation period $P$ was treated as a free parameter and assigned an empirical Gaussian prior. Rather than relying on literature values, the prior's central value was derived directly from the observational data using a generalized multiband periodogram. This periodogram explicitly accounted for the varying passband offsets and phase angle dependencies via a regularized linear regression framework. The frequency corresponding to the maximum spectral power was identified, and the resulting period was adopted as the mean of the Gaussian prior. To encompass potential local structural variations in the posterior distribution while preventing unbounded wandering, this prior was assigned a standard deviation of $0.005$~hours. The resulting best-fit rotational model, phase-folded over the converged period alongside the observational data, is presented in the left panel of Fig.~\ref{fig:multiband_diag}.

\begin{figure*}[ht]
	\centering
	\includegraphics[width=0.75\textwidth]{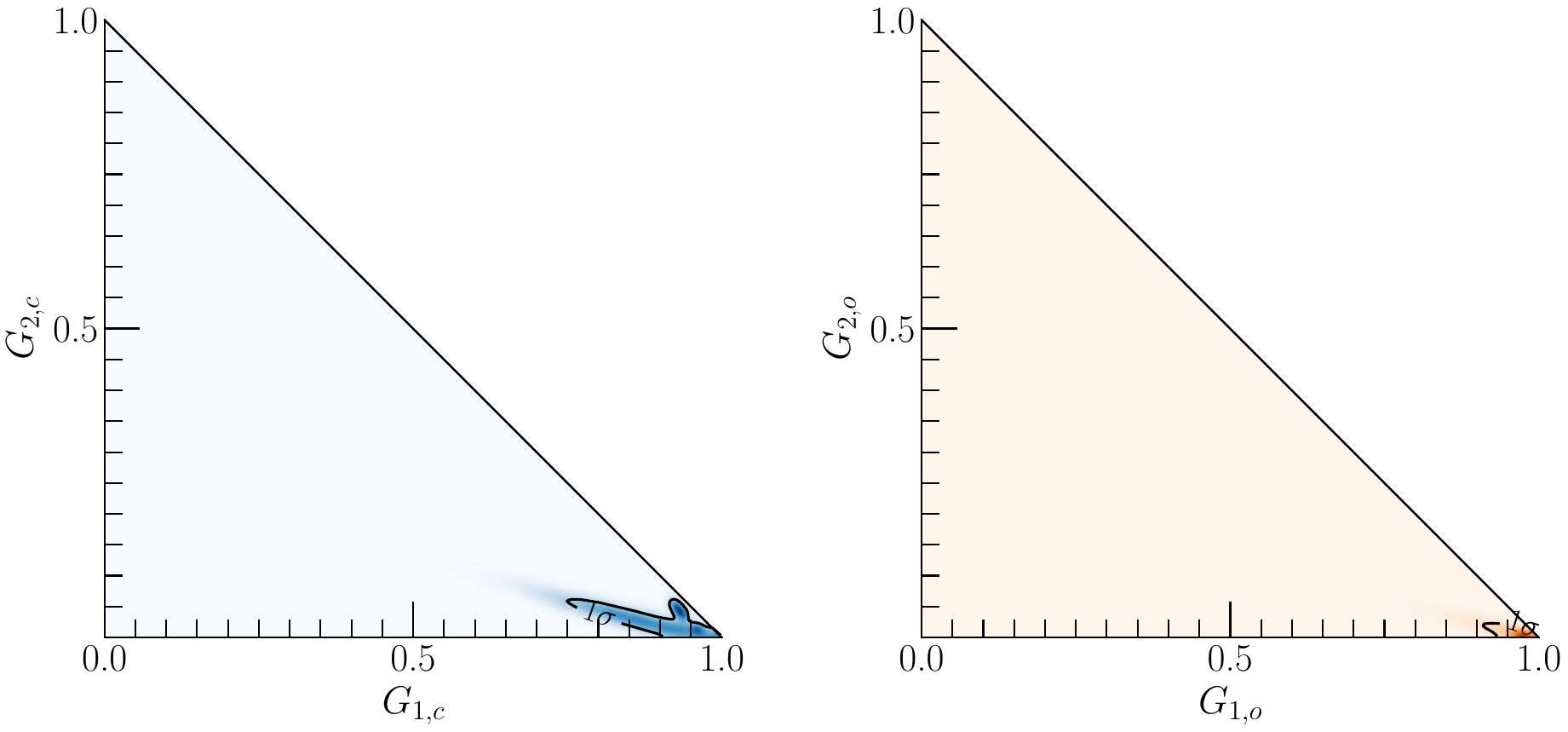}
	\caption{Distribution of the fitted $G_1$ and $G_2$ parameters for the ATLAS $c$- and $o$-band data (left and right panels, respectively).}\label{fig:g1g2_dist}
\end{figure*}

For the phase curve construction (Fig.~\ref{fig:multiband_diag}, right panel), we dynamically bin the data based on the phase angle ($\alpha$), applying an opposition cutoff at $\alpha = 10^\circ$. The applied bin width ($\Delta\alpha$) is defined as:
\begin{equation}
	\Delta\alpha=\begin{cases}\Delta\alpha_\text{fine}=\frac{h}{3}&\text{for }\alpha<10^\circ\\\Delta\alpha_\text{coarse}=h&\text{for }\alpha\geq10^\circ\end{cases},\label{eq:binned_alpha}
\end{equation}
where $h$ is the nominal base bin width. We set a base width of $h = 5^\circ$ and a minimum of $N = 20$ data points per bin as default values. These are adjusted according to the data density in each filter, ensuring we retain a minimum of three fine bins ($N_\text{bins} \geq 3$) in the opposition region. This maintains an optimal balance between the scarcity of observations at low phase angles and the goodness of fit required for the non-linear portion of the phase curve. During the simultaneous MCMC fine-tuning, the binned data for each filter were evaluated separately using the $H, G_1, G_2$ phase function implemented in the \texttt{sbpy} package \citep{Mommert2019}. This independent filter evaluation is necessary because asteroid phase curves exhibit phase reddening, which leads to passband-dependent $G_1$ and $G_2$ parameters \citep{Mahlke2021}. The final derived parameters and their posterior distributions are presented in Table~\ref{tab:hg1g2_results}, and the $G_1$ and $G_2$ distribution plots for each band are shown in Fig.~\ref{fig:g1g2_dist}.

\begin{figure*}[ht]
	\centering
	\includegraphics[width=0.75\textwidth]{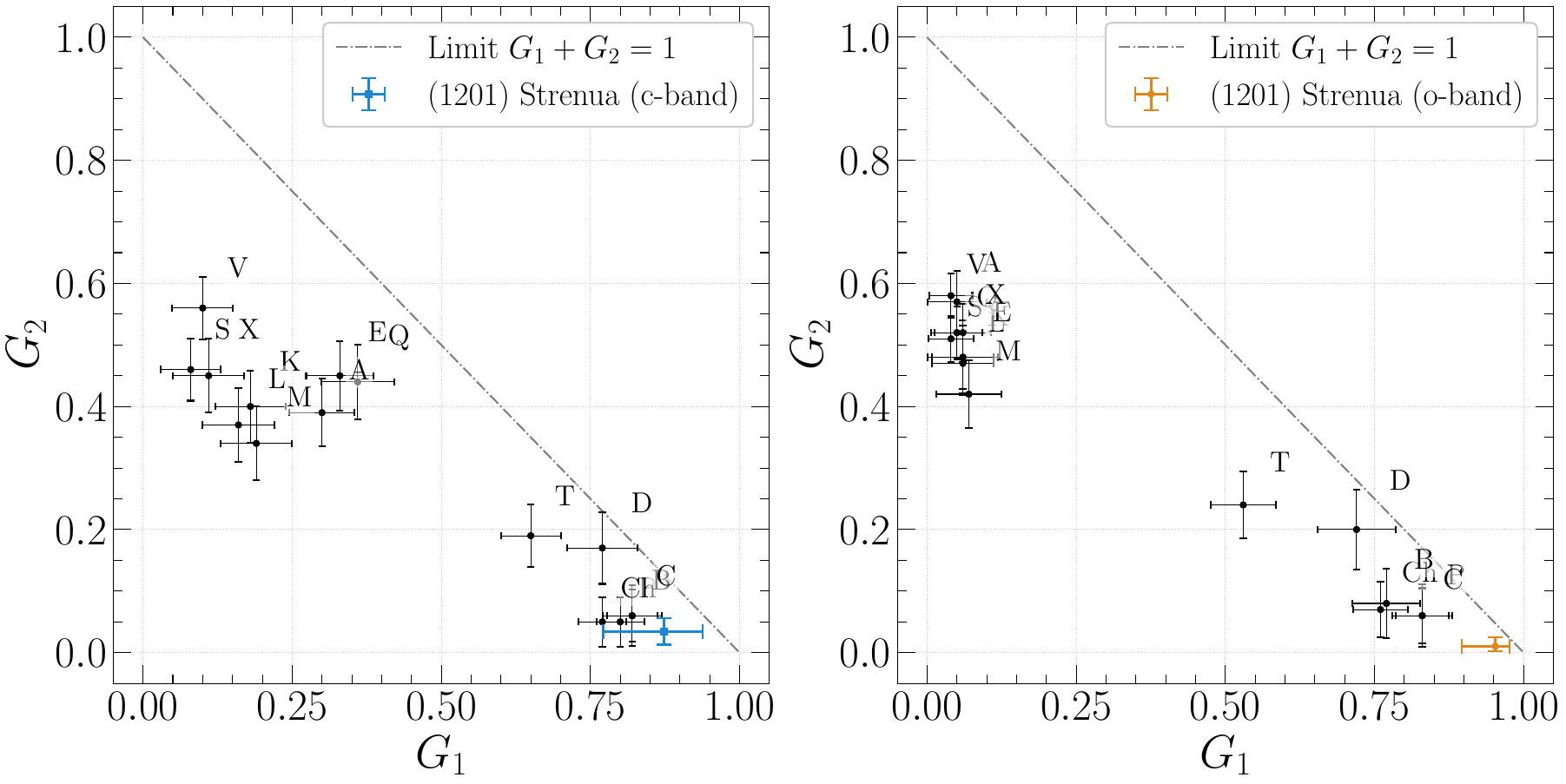}
	\caption{The derived $G_1$ and $G_2$ parameters of (1201) Strenua in each photometric band, shown in the context of the $G_1$--$G_2$ distributions of several asteroid taxonomic classes reported by \citet{Mahlke2021}.}\label{fig:g1g2_taxonomy}
\end{figure*}

\begin{table*}[ht]
	\centering
	\caption{Phase curve fitting results for the ATLAS $c$- and $o$-band. The table presents the derived absolute magnitudes ($H$) and phase parameters ($G_1, G_2$) with their uncertainties, alongside the number of photometric data points ($N$) and the phase angle ($\alpha$) coverage used for the fit.}\label{tab:hg1g2_results}
	\renewcommand{\arraystretch}{1.5}
	\begin{tabular}{
			crcccc
		}
		\hline\hline
		\textbf{Band} & \bm{$N$} & \bm{$\alpha$} & \bm{$H$} & \bm{$G_1$} & \bm{$G_2$} \\
		\hline
		$c$ & 507 & $1.97^\circ$ to $22.78^\circ$ & $11.50^{+0.05}_{-0.05}$ & $0.81^{+0.08}_{-0.08}$ & $0.04^{+0.02}_{-0.02}$ \\
		$o$ & 1879 & $1.69^\circ$ to $22.98^\circ$ & $11.29^{+0.02}_{-0.03}$ & $0.94^{+0.03}_{-0.05}$ & $0.01^{+0.01}_{-0.01}$ \\
		\hline
	\end{tabular}
\end{table*}   

\subsection{Geometric Albedo and Taxonomy Determination}
Previous large-scale analyses of ATLAS dual-band photometry by \citet{Mahlke2021} demonstrated that these specific phase curve parameters serve as a robust proxy for taxonomic classification when detailed spectroscopy is unavailable. As illustrated in Fig.~\ref{fig:g1g2_taxonomy}, projecting our derived $G_1$ and $G_2$ values against these established distributions maps Strenua securely into the domain predominantly occupied by low-albedo, primitive taxonomic classes \citep{Penttila2016, Mahlke2021}. The combination of high $G_1$ parameters ($>0.8$) and near-zero $G_2$ parameters ($<0.05$) points to a highly backscattering surface characterized by a steep linear phase slope and a prominent opposition surge. Such photometric signatures are typical of highly porous, uncompacted regoliths where single-particle scattering dominates over multiple scattering, as is commonly observed in dark carbonaceous or organic-rich bodies \citep{Belskaya2000, Shevchenko2016}.

To translate these photometric insights into a definitive geometric albedo, we utilize the equivalent diameter derived from our multi-chord occultation observations (Section~\ref{sec:chord}; cf. Table~\ref{tab:strenua_parameters}). To properly apply Equation~\ref{eq:albedo_absmag}, the absolute magnitudes obtained in the ATLAS $c$- and $o$-bands (Table~\ref{tab:hg1g2_results}) were first converted to the Johnson $V$-band absolute magnitude, $H_\mathrm{V}$, using the empirical transformation provided by \citet{Tonry2025}:
\begin{equation}
	\left(H_\mathrm{V}-H_o\right) = -0.007 + 0.884\left(H_c-H_o\right)\label{eq:hv_conversion}
\end{equation}

We derived an ATLAS pseudo-color index of $H_c - H_o = 0.21 \pm 0.06$ mag to further constrain the surface composition. This positive color index indicates a higher reflectance in the longer-wavelength $o$-band compared to the $c$-band, signifying a moderately red spectral slope in the visible wavelength regime. Concurrently, linking our synthesized $H_\mathrm{V}$ with the occultation diameter yields an extremely low geometric visual albedo of $p_\mathrm{V} = 0.029 \pm 0.003$. This value falls significantly below the typical albedo ranges for silicaceous S-complex ($p_\mathrm{V} \approx 0.15$--$0.25$) or even standard carbonaceous C-complex asteroids ($p_\mathrm{V} \approx 0.04$--$0.10$). Instead, it aligns strongly with the most primitive, opaque populations in the asteroid belt \citep{Mainzer2011, Usui2011}.

Historically, featureless spectral surveys have often associated (1201) Strenua with the legacy X-complex taxonomy. However, the X-complex is well-known for its severe taxonomic degeneracy, grouping highly reflective E-types, moderate-albedo M-types, and dark P-types based purely on spectral shape without resolving their vastly different surface compositions \citep{Tholen1984, DeMeo2009}. To resolve such ambiguities, the modern taxonomic framework outlined by \cite{Mahlke2022} reintroduces the visual geometric albedo alongside visible and near-infrared spectral slopes to probabilistically assign classes. Under the Mahlke et al. system, the traditional X-complex degeneracy is explicitly broken; the P-class is structurally grouped back with primitive C-complex bodies based on their low albedos, and a new Z-class is introduced for ultra-red, dark objects. By applying this new taxonomic paradigm to our results, the E- and M-type classifications for Strenua are definitively eliminated. The extremely low geometric albedo ($p_\mathrm{V} = 0.029$) combined with the moderately red visible spectral slope places (1201) Strenua firmly within the darkest primitive classes. While standard P-types in the Mahlke taxonomy typically occupy the $0.04 \le p_\mathrm{V} \le 0.07$ regime, our derived albedo is notably lower, pushing Strenua into a domain characteristic of the darkest C-complex or D-type asteroids (where $p_\mathrm{V} < 0.12$). By synthesizing the extreme low albedo, the color index, and the phase parameters that cluster tightly in the dark-asteroid locus, we robustly characterize (1201) Strenua as a highly opaque, primitive object. This classification conforms completely with the \cite{Mahlke2022} taxonomic re-evaluation, confirming that the surface composition is rich in organics, carbonaceous materials, and anhydrous silicates typical of the primordial populations residing in the outer main belt.


\section{Conclusions and Future Work}
\label{sec:conclusions}

We presented a comprehensive physical characterization of the main-belt asteroid (1201) Strenua by synergizing high-precision stellar occultation timing with archival ATLAS photometry. By employing a rigorous Bayesian forward-modeling framework that accounts for point-source Fresnel diffraction and finite stellar disk effects, we successfully extracted the instantaneous physical dimensions of the occulting body. Specifically, we derived apparent semi-axes of $a' = 27.49^{+3.90}_{-1.65}$ km and $b' = 14.33^{+2.37}_{-2.69}$ km, corresponding to a full cross-sectional profile of $54.98^{+7.80}_{-3.30} \times 28.66^{+4.74}_{-5.38}$ km and an area-equivalent diameter of $D_\text{eq} = 39.86^{+2.06}_{-1.66}$ km. Combined with a derived position angle of $95.3^{+4.9}_{-2.7}$ deg, these parameters provide a highly precise astrometric and geometric anchor for Strenua at the epoch of observation.

Furthermore, the integration of ATLAS $c$- and $o$-band photometry enabled robust phase curve modeling. We derived band-specific absolute magnitudes of $H_c = 11.50 \pm 0.05$ mag and $H_o = 11.29^{+0.02}_{-0.03}$ mag, alongside the phase function parameters $G_{1,c} = 0.81 \pm 0.08$ and $G_{2,c} = 0.04 \pm 0.02$ for the $c$ band, and $G_{1,o} = 0.94^{+0.03}_{-0.05}$ and $G_{2,o} = 0.01 \pm 0.01$ for the $o$ band. From these ATLAS measurements, we calculated the standard V-band absolute magnitude, yielding $H_\mathrm{V} = 11.46 \pm 0.05$ mag. Utilizing the size-magnitude relationship, we combined this $H_\mathrm{V}$ value with our occultation-derived equivalent diameter to tightly constrain Strenua's geometric albedo to $p_\mathrm{V} = 0.029 \pm 0.003$. By projecting our derived $G_1$ and $G_2$ values against established distributions, we confidently mapped Strenua into the P-class taxonomic cluster, establishing a coherent physical link between its geometrically derived size and its surface scattering properties.

Taken together, these results illustrate the value of combining geometrically absolute occultation sizing with phase-curve-based taxonomy: the two techniques are derived from entirely independent observables (chord timing versus disk-integrated flux) yet converge on a mutually consistent picture of a dark, backscattering-dominated surface. This agreement is notable because it resolves, for this individual object, the long-standing degeneracy of the legacy X-complex classification without requiring new spectroscopic input. The resulting albedo, $p_\mathrm{V} = 0.029 \pm 0.003$, is low even relative to typical P-class values, suggesting that Strenua's surface may sit toward the darkest, most opaque end of the primitive population rather than representing a typical P-type member, a distinction that future compositional work should test directly. 

Despite these advancements, the inherent limitations of single-epoch and unresolved observations leave room for further refinement. While our current occultation data yields a precise 2D cross-sectional profile, it represents only a single rotational phase. Future multi-chord occultation campaigns are critical, as denser chord coverage will allow for the direct derivation of a complete 3D shape model and a more accurate volume-equivalent diameter. These geometric constraints should ideally be coupled with light-curve inversion techniques applied to a broader dataset of dense photometry. Doing so will enable the derivation of a robust spin-pole solution, resolving ambiguities in our current projection and allowing for the reconstruction of a global convex shape model.

In addition to geometric refinements, further investigation into Strenua's surface properties is highly warranted. The derived phase curve parameters could be improved with targeted, high-cadence photometric observations at very low phase angles, as constraining the non-linear opposition surge is vital for refining both the absolute magnitude and the resulting geometric albedo. Finally, although our $G_1$ and $G_2$ parameter mapping provides a strong taxonomic indicator, near-infrared and visible spectroscopy are ultimately necessary to definitively confirm Strenua's surface composition. Dedicated spectroscopic follow-up will validate the albedo-taxonomy consistency derived in this study and confidently place the asteroid within a broader compositional and evolutionary context.


\begin{acknowledgments}
We extend our deepest gratitude to the students, amateur astronomers, and volunteers who participated in this campaign. The successful deployment of mobile stations highlights the critical value of Pro-Am collaborations in capturing multi-chord asteroidal occultations. We particularly thank Rizki Sidik (FST Undana), Chornelis J. B. Anin (Geography Education Study Program Undana), and Audy R. F. Liukae (Pecinta Langit Timor) for their valuable contributions to the observation campaign. A.T.P.J. extends his gratitude to Dr. rer. nat. Hesti R. T. Wulandari, Director of Bosscha Observatory, alongside Luthfiandari (Bosscha Observatory), Yatny Yulianty (Bosscha Observatory), Fauzia La Fatsa (Bosscha Observatory), Fatimah Zahra (Bosscha Observatory), Amandaru T. Mandala (Master's Program in Astronomy ITB, Bosscha Observatory), Mochamad Irfan (Bosscha Observatory), and Rudi Hamdani (Bosscha Observatory), for their invaluable support during the workshops, colloquia, and pre-campaign preparations. We also gratefully acknowledge the support provided by Leids Kerkhoven-Bosscha Fonds (LKBF). Finally, A.T.P.J. dedicates this work to the memory of the late Dr. Andrea Richichi (INAF--Osservatorio Astrofisico di Arcetri, Italy), gratefully acknowledging his enduring influence and for first introducing him to the field of occultation research.
\end{acknowledgments}

\appendix

\section{STRENUA-26 Contributors}
\label{sec:appendix_contributors}

The STRENUA-26 Contributors comprise the observers, station principal investigators, and volunteers who successfully deployed monitoring equipment during the 26 April 2026 occultation campaign. The complete observational circumstances, instrumental configurations, and event timings for each station are detailed in Table~\ref{tab:list_all_stations}.

\startlongtable
\begin{deluxetable}{
		>{\raggedright\arraybackslash}p{2.5cm}
		>{\raggedright\arraybackslash}p{2.1cm}
		r
		r
		>{\raggedright\arraybackslash}p{1.75cm}
		r
		c
		c
		>{\raggedright\arraybackslash}p{2.95cm}
	}
	\setlength{\LTcapwidth}{0.9\textwidth}
	\renewcommand{\arraystretch}{0.90}
	\tabletypesize{\scriptsize}
	\setlength{\tabcolsep}{2pt}
	\tablecaption{Observational circumstances and outcomes for the 26 April 2026 stellar occultation by (1201) Strenua.\label{tab:list_all_stations}}
	\tablewidth{\textwidth}
	\tablehead{
		\colhead{Institution} &
		\colhead{Station} &
		\colhead{Lon. / Lat.} &
		\colhead{Aperture} &
		\colhead{Camera} &
		\colhead{Exposure} &
		\colhead{Time} &
		\colhead{Status} &
		\colhead{Observers} \\
		\colhead{} &
		\colhead{} &
		\colhead{Elev. (deg / m)} &
		\colhead{(mm)} &
		\colhead{} &
		\colhead{(s)} &
		\colhead{Source} &
		\colhead{} &
		\colhead{}
	}
	\startdata
	Observatorium Bosscha & BOSSCHA-2 & \makecell[r]{123$^\circ$ 27' 52.53'' E \\ 10$^\circ$ 20' 16.23'' S \\ 0} & 106 & QHY174M GPS & $0.030$ & GPS & P & M. Yusuf, C. J. B. Anin, A. R. F. S. Liukae \\
	Penjelajah Langit & PL-2 & \makecell[r]{110$^\circ$ 31' 48.00'' E \\ 07$^\circ$ 14' 49.20'' S \\ 403} & 61 & SVBONY SV905C & $0.150$ & NTP & P & M. I. B. Hersulistyo, F. R. Atmaja, N. R. Sholihah, D. Ardiana, Merdalena \\
	Planetarium dan Observatorium UIN Walisongo Semarang & UINWS-2 & \makecell[r]{110$^\circ$ 17' 49.20'' E \\ 07$^\circ$ 04' 55.20'' S \\ 259} & 80 & QHY5-IIM & $0.055$ & NTP & P & M. S. Fadhel, M. Adam, M. Saifullah, F. Farrukh, M. Inayah, R. Syam, K. Anaam, M. Ulil A., K. Umam, I. Ma \\
	Arisan Astro; Observatorium Azcotico, Kupang & ZUL & \makecell[r]{123$^\circ$ 27' 54.00'' E \\ 10$^\circ$ 20' 16.80'' S \\ 0} & 70 & ZWO ASI294MC-Pro & $0.031$ & NTP & P & Zulkarnain \\
	Observatorium Astronomi Itera Lampung (OAIL ITERA) & ITERA-6 & \makecell[r]{105$^\circ$ 35' 22.05'' E \\ 05$^\circ$ 44' 28.64'' S \\ 22} & 205 & QHY174M GPS & 0.005 & GPS & N & A. Oktaviandra, W. Sepindarto, M. P. Mahayu, Septi, G. Maharani \\
	Observatorium Astronomi Itera Lampung (OAIL ITERA) & ITERA-5 & \makecell[r]{105$^\circ$ 35' 22.08'' E \\ 05$^\circ$ 44' 28.50'' S \\ 22} & 80 & ZWO ASI533MM & $0.100$ & NTP & N & A. R. D. Putri, A. Z. Rozzykin, A. B. A. Pelupessy \\
	Observatorium Bosscha & BOSSCHA-1 & \makecell[r]{107$^\circ$ 37' 0.22'' E \\ 06$^\circ$ 49' 28.71'' S \\ 1310} & 279.4 & ZWO ASI290M & $0.025$ & NTP & N & A. T. P. Jatmiko, D. Mandey, B. Dermawan, A. T. Mandala, K. T. Islamy, M. F. A. Triady, P. W. Premadi \\
	Observatorium Astronomi Itera Lampung (OAIL ITERA) & ITERA-1 & \makecell[r]{105$^\circ$ 18' 50.88'' E \\ 05$^\circ$ 21' 37.42'' S \\ 130} & 254 & ZWO ASI2600MM Pro & $0.040$ & NTP & N & A. A. Yusuf, M. D. Nuramalina, H. R. Aulia, D. S. N. Azizah, A. S. Qathrunnada, L. N. W. Yumna \\
	Prodi Sains Atmosfer dan Keplanetan ITERA, OAIL ITERA & ITERA-2 & \makecell[r]{105$^\circ$ 18' 50.94'' E \\ 05$^\circ$ 21' 37.57'' S \\ 130} & 200 & ZWO ASI174MM Mini & $0.008$ & NTP & N & A. N. I. Putri, N. A. Ferdian, M. Z. Alwan, M. B. Aflah, A. L. Sipayung, F. Ainurrohmah, Juna, Raudina \\
	Prodi Sains Atmosfer dan Keplanetan ITERA; OAIL ITERA & ITERA-3 & \makecell[r]{105$^\circ$ 18' 51.02'' E \\ 05$^\circ$ 21' 37.37'' S \\ 130} & 70 & ZWO ASI678MC & $0.050$ & NTP & N & M. F. Junaidi, S. Narisma, R. Legowo, Z. R. Usbaya, Wafa, Naufal \\
	Observatorium Bosscha & BOSSCHA-3 & \makecell[r]{107$^\circ$ 37' 15.03'' E \\ 06$^\circ$ 48' 14.74'' S \\ 1299} & 66 & ZWO ASI290MM & $0.030$ & NTP & N & S. Fatkhusalma, T. Perhati, F. Zahra \\
	Planetarium dan Observatorium UIN Walisongo Semarang & UINWS-1 & \makecell[r]{110$^\circ$ 20' 52.80'' E \\ 06$^\circ$ 59' 31.20'' S \\ 109} & 61 & QHY5-IIM & $0.050$ & NTP & N & F. A. Fuscha, T. Amanah P., M. Z. Aziz, M. H. F. Hadil W., W. Cosara, M. T. Langit, S. Ramdani, Farhan, Naufal \\
	Observatorium Pondok Tahfidz Yanbu'ul Qur'an Menawan & MENAWAN & \makecell[r]{110$^\circ$ 51' 03.60'' E \\ 06$^\circ$ 41' 52.80'' S \\ 282} & 98 & ZWO ASI290MM & $0.020$ & NTP & N & N. Sidqon \\
	N/A & BIMANASUTION & \makecell[r]{106$^\circ$ 49' 26.40'' E \\ 06$^\circ$ 30' 39.60'' S \\ 156} & 203.2 & ZWO ASI120MM & -- & NTP & NDR & B. Nasution \\
	HAAJ & HAAJ-3 & \makecell[r]{106$^\circ$ 44' 32.70'' E \\ 06$^\circ$ 29' 07.24'' S \\ 270} & 130 & QHY600PH SBFL & -- & NTP & NDR & Aripudin \\
	Malang Photo Club (MPC) & KEC9SO & \makecell[r]{112$^\circ$ 38' 16.80'' E \\ 07$^\circ$ 56' 52.80'' S \\ 500} & 150 & ZWO ASI183MM & -- & NTP & NDR & P. Nugroho \\
	MTs Muhammadiyah Malang & GVR-MLG & \makecell[r]{112$^\circ$ 37' 22.80'' E \\ 07$^\circ$ 54' 54.00'' S \\ 490} & 54 & Canon EOS 40D\tablenotemark{a} & -- & NTP & NDR & Ichi \\
	Ngawi Astronomy Club & BKI & \makecell[r]{111$^\circ$ 27' 39.60'' E \\ 07$^\circ$ 22' 01.20'' S \\ 210} & 61 & QHY5L Mono & -- & NTP & NDR & Rhody A. Saputro \\
	OIF UMSU & OIFUMSU & \makecell[r]{98$^\circ$ 43' 15.60'' E \\ 03$^\circ$ 34' 55.20'' N \\ 32} & 71 & ZWO ASI178MM & -- & NTP & NDR & H. Putraga \\
	SMAN 1 Sumedang & BERRELGADAMA & \makecell[r]{107$^\circ$ 54' 46.80'' E \\ 06$^\circ$ 50' 13.20'' S \\ 474} & 60 & Smartphone Redmi 13C\tablenotemark{a} & -- & NTP & NDR & B. Gadama \\
	Starlight Explorer & STARLIGHT\-EXPLORER-01 & \makecell[r]{106$^\circ$ 40' 48.00'' E \\ 06$^\circ$ 29' 38.40'' S \\ 125} & 152.4 & ZWO ASI120MM-S & -- & NTP & NDR & E. Kristian \\
	UIN Sunan Ampel Surabaya & SHAULA\-OBSERVATORY & \makecell[r]{107$^\circ$ 35' 44.15'' E \\ 06$^\circ$ 50' 45.02'' S \\ 1000} & 80 & SVBONY SV465CC & -- & NTP & NDR & N. Sopwan \\
	Pusat Studi Astronomi UAD & PASTRON UAD-3 & \makecell[r]{110$^\circ$ 22' 58.80'' E \\ 07$^\circ$ 49' 58.80'' S \\ 41} & 80 & QHY 183C & -- & NTP & NDR & Y. H. Perdana, A. Fauzi, T. S. B. Sleman \\
	Pusat Studi Astronomi UAD & PASTRON UAD-1 & \makecell[r]{110$^\circ$ 10' 22.80'' E \\ 07$^\circ$ 21' 50.40'' S \\ 676} & 750 & Skyris 274M & -- & NTP & NDR & M. Maulidan, A. L. Calista, M. Azazkiya \\
	Program Studi Ilmu Falak, IAIN Curup & IAINCURUP & \makecell[r]{102$^\circ$ 31' 26.40'' E \\ 03$^\circ$ 27' 00.00'' S \\ 600} & 80 & ZWO ASI585MC & -- & NTP & NDR & R. Soderi, A. A. Putra, K. U. Khudhari \\
	Observatorium Bosscha & BOSSCHA-4 & \makecell[r]{107$^\circ$ 38' 47.00'' E \\ 06$^\circ$ 43' 47.00'' S \\ 1230} & 66 & ZWO ASI120MM & -- & NTP & NDR & M. Z. Aprianizar, D. G. Ramadhan, R. Hamdani \\
	Pusat Studi Astronomi UAD & PASTRON UAD-2 & \makecell[r]{110$^\circ$ 13' 48.00'' E \\ 07$^\circ$ 25' 48.00'' S \\ 300} & 71 & ZWO ASI678MC & -- & NTP & NDR & Y. Pramudya, E. Sriruenroeng, S. Wahyuni \\
	Universitas Pendidikan Indonesia & IMAHNOONG-2 & \makecell[r]{107$^\circ$ 37' 06.00'' E \\ 06$^\circ$ 49' 56.00'' S \\ 1300} & 150 & ZWO 780MM-S & -- & NTP & NDR & J. A. Utama, F. Apryandi, D. S. Hardiyanti, E. F. Uyun, H. Z. A. Hamzah, S. A. Immanuel, Z. A. Yusuf \\
	UIN Sultan Maulana Hasanuddin Banten & UINSMHB & \makecell[r]{106$^\circ$ 09' 18.00'' E \\ 06$^\circ$ 11' 02.40'' S \\ 95} & 108 & QHY183C & -- & NTP & NDR & S. Pramono, H. Ahdi \\
	Lingkaran Studi Ilmu Hisab Rukyat AL-KAUKABA Lamongan & ALKAUKABA & \makecell[r]{112$^\circ$ 25' 26.40'' E \\ 07$^\circ$ 00' 43.20'' S \\ 38} & 90 & SVBONY SC715C & -- & NTP & NDR & M. K. Anam, M. S. Khoir, M. R. R. Al-Ahady \\
	Penjelajah Langit & PL-1 & \makecell[r]{110$^\circ$ 23' 27.60'' E \\ 07$^\circ$ 16' 51.60'' S \\ 207} & 60 & ZWO ASI120MM & -- & NTP & NDR & E. H. Gunawan, M. C. S. Sari, M. A. M. Hayati, S. Pambudi, A. Cahyani \\
	Badan Riset dan Inovasi Nasional & BRIN & \makecell[r]{106$^\circ$ 42' 00.00'' E \\ 06$^\circ$ 18' 00.00'' S \\ 45} & 152 & Sony IMX678 Color\tablenotemark{b} & -- & NTP & NDR & T. Djamaluddin, R. A. Pandapotan \\
	Assalaam Observatory & CASA & \makecell[r]{110$^\circ$ 46' 15.60'' E \\ 07$^\circ$ 33' 10.80'' S \\ 175} & 50 & Sony IMX462 Color\tablenotemark{c} & -- & NTP & NDR & A. S. Riadi, T. R. Wulandari \\
	IMAHNOONG & IMAHNOONG-1 & \makecell[r]{107$^\circ$ 37' 08.40'' E \\ 06$^\circ$ 49' 55.20'' S \\ 1300} & 150 & ZWO 780MM-S & -- & NTP & NDR & H. Setyanto \\
	Planetarium Ranting Sewu & RANTINGSEWU & \makecell[r]{112$^\circ$ 41' 09.60'' E \\ 07$^\circ$ 40' 37.20'' S \\ 250} & 50 & Sony IMX462 Color\tablenotemark{c} & -- & NTP & NDR & H. Nasiruddin \\
	Tirta Bening Observatory & TIRTABENINGOBS & \makecell[r]{112$^\circ$ 40' 12.00'' E \\ 07$^\circ$ 42' 03.60'' S \\ 440} & 102 & QHY 178M & -- & NTP & NDR & J. Anderson \\
	WRW Observatory & ROBIN & \makecell[r]{105$^\circ$ 50' 02.40'' E \\ 06$^\circ$ 13' 04.80'' S \\ 26} & 130 & QHY174M GPS & -- & GPS & NDR & Robin \\
	AstroBali & ASTROBALI & \makecell[r]{115$^\circ$ 21' 10.80'' E \\ 08$^\circ$ 27' 39.60'' S \\ 408} & 150 & SVBONY 205 & -- & NTP & NDR & R. I. Yafityan \\
	Universitas Islam Negeri Mataram & UINMATARAM & \makecell[r]{116$^\circ$ 06' 03.60'' E \\ 08$^\circ$ 36' 36.00'' S \\ 13} & 125 & ZWO ASI174MM & -- & NTP & NDR & M. D. Firdaus \\
	Griya Antariksa & ANTARIKSA & \makecell[r]{110$^\circ$ 23' 42.00'' E \\ 07$^\circ$ 45' 36.00'' S \\ 152} & 80 & ZWO ASI178 & -- & NTP & NDR & M. Arkanuddin \\
	\enddata
	\tablenotetext{a}{Station deployed as part of the campaign's educational outreach and citizen science initiative.}
	\tablenotetext{b}{Integrated detector for the Celestron Origin Mark II smart telescope.}
	\tablenotetext{c}{Integrated detector for the ZWO Seestar S50 smart telescope.}
	\tablecomments{The participating observers listed in
		this table collectively constitute the STRENUA-26 Contributors. Station status is denoted as positive (P), negative (N), or no data recorded (NDR) for sites where	observations were precluded by technical anomalies or adverse weather. For both smart telescope systems ($^\text{b,c}$), absolute timestamping is mediated via a mobile application interface, introducing uncalibrated software and network latencies that limit the reliability of standard NTP synchronization for high-precision occultation timing.}
\end{deluxetable}

\bibliography{references}{}
\bibliographystyle{aasjournalv7.1}



\end{document}